\documentclass[preprint]{elsarticle}
\usepackage{amsmath}
\usepackage{color}
\newcommand{\be}{\begin{equation}}\newcommand{\ee}{\end{equation}}
\newcommand{\bea}{\begin{eqnarray}}\newcommand{\eea}{\end{eqnarray}}
\newcommand{\brr}{\begin{array}}\newcommand{\err}{\end{array}}
\newcommand{\bit}{\begin{itemize}}\newcommand{\eit}{\end{itemize}}
\newcommand{\ben}{\begin{enumerate}}\newcommand{\een}{\end{enumerate}}

\newcommand{\bbm}{\begin{bmatrix}}\newcommand{\ebm}{\end{bmatrix}}
\newcommand{\ba}{\begin{array}}
\newcommand{\ea}{\end{array}}
\newcommand{\G}{\textbf}

\newtheorem{mydef}{Definition}
\newtheorem{Lemma}{Lemma}
\newtheorem{theorem}{Theorem}
\newcommand{\bd}{\begin{mydef}} \newcommand{\ed}{\end{mydef}}
\newcommand{\bthe}{\begin{theorem}} \newcommand{\ethe}{\end{theorem}}
\newcommand{\ble}{\begin{Lemma}} \newcommand{\ele}{\end{Lemma}}

\newcommand{\dr}{\mathrm{d}}

\definecolor{darkred}{rgb}{.8,0,0}

\definecolor{darkblue}{rgb}{0,0,.7}

\def\ha{\frac{1}{2}}

\def\intk{\int \!\!\mathrm{d}^3 {\G k}}

\def\lan{\langle}
\def\lf{\left}

\def\non{\nonumber}\def\pa{\partial}\def\ran{\rangle}

\def\ri{\right}
\def\al{\alpha}\def\bt{\beta}
\def\de{\delta}\def\ep{\epsilon}
\def\te{\vartheta}

\def\om{\omega}

\def\1{{_{1}}}\def\2{{_{2}}}
\newcommand{\ide}{1\hspace{-1mm}{\rm I}}

\def\noHe0{:\;\!\!\;\!\!:H_e(0):\;\!\!\;\!\!:}
\def\noHm0{:\;\!\!\;\!\!:H_\mu(0):\;\!\!\;\!\!:}

\def\lan{\langle}
\def\lf{\left}

\def\non{\nonumber}
\def\pa{\partial}\def\ran{\rangle}

\def\ri{\right}

\def\al{\alpha}\def\bt{\beta}
\def\de{\delta}
\def\ep{\epsilon}\def\te{\theta}

\def\om{\omega}

\def\1{{_{1}}}\def\2{{_{2}}}

\begin{document}

\title{Functional integrals and inequivalent representations in Quantum Field Theory}

\author[su]{M~Blasone}
\ead{blasone@sa.infn.it}

\author[cvut]{P~Jizba}
\ead{p.jizba@fjfi.cvut.cz}

\author[su]{L~Smaldone}
\ead{lsmaldone@sa.infn.it}

\address[su]{Dipartimento di Fisica, Universit\`a di Salerno, Via Giovanni Paolo II, 132 84084 Fisciano, Italy \& INFN Sezione di Napoli, Gruppo collegato di Salerno, Italy}
\address[cvut]{FNSPE, Czech Technical University in Prague, B\v{r}ehov\'{a} 7, 115 19 Praha 1, Czech Republic}

\begin{abstract}
We discuss canonical transformations in Quantum
Field Theory in the framework of the functional-integral approach.
In contrast with ordinary Quantum Mechanics, canonical transformations in Quantum Field Theory are mathematically more subtle due to the existence of unitarily inequivalent representations of canonical commutation relations. When one works with functional integrals, it is not immediately clear how this algebraic feature manifests itself in the formalism. Here we attack this issue by considering the canonical transformations in the context of coherent-state functional integrals.
Specifically, in the case of linear canonical transformations, we derive the general  functional-integral representations for both transition amplitude and partition function phrased in terms of new canonical variables. By means of this, we show how in the infinite-volume limit the canonical transformations induce a transition from one representation of canonical commutation relations to another one and under what conditions the representations are unitarily inequivalent.
We also consider the partition function and derive the energy gap between statistical systems described in two different representations which,
among others, allows to establish a connection with continuous phase transitions.  We illustrate the inner workings of the outlined mechanism by discussing two prototypical systems:
the van Hove model and the Bogoliubov model of weakly interacting Bose gas.
\end{abstract}

\begin{keyword}
Theory of quantized fields \sep Functional integrals \sep Canonical transformations
\sep Coherent states
\PACS 03.70.+k \sep 02.90.+p \sep  03.65.Db \sep 03.65.Sq	
\end{keyword}

\maketitle
\section{Introduction}

Soon after the formulation of Quantum Field Theory (QFT), it became clear that
this is not a straightforward extension of Quantum Mechanics (QM) to field-theoretic systems~\cite{schweber,ItzZub}. The issue at stake is indeed  more delicate than that: QFT deals with systems having an infinite number of degrees of freedom and hence the Stone--von Neumann uniqueness theorem~\cite{Neu}, establishing the unitary equivalence among different representations of the canonical commutation relations (CCR), is not applicable. Consequently, there exists in QFT infinitely many unitarily inequivalent representations of the field algebra~\cite{bratt}, or equivalently, for a given dynamics, there is an infinite number of (inequivalent) physical realizations. It should be stressed that inequivalent quantum vacua may also stem from a non-trivial topology of a classical configuration space, as, e.g., in the Aharonov--Bohm effect~\cite{Kastrup}. In this
case the Stone--von Neumann theorem does not apply, irrespective of the number of degrees of freedom. These types of inequivalent vacua will not be considered here.

An important, and not often appreciated, aspect of QFT is that the interacting (Heisenberg) fields do not have a unique representation in terms of the asymptotic (quasi-)particle fields, i.e., fields which directly act on the Fock space and whose elementary excitations
are directly susceptible of experimental detections. In fact, the functional relation between the asymptotic fields and Heisenberg fields, known as {\em Haag's map}, is only a weak operatorial relation, i.e., it is valid only for matrix elements of the operators constructed with respect to the base states of the Hilbert space of asymptotic fields. All these facts are
of course well known, and indeed important phenomena such as the spontaneous symmetry breakdown~\cite{UMZ2,Sewell,BJV}, renormalization~\cite{BJV,UTK,Haag}, Hawking's black hole radiation~\cite{Haag2,Hawking, MaSoVi}, quantization on curved backgrounds~\cite{Birrell} or quantization of dissipative systems~\cite{BJV,CeRaVi}, can be successfully addressed with QFT only because one can invoke the concept of inequivalent representations.

One might also mention that recently the essential r\^{o}le of inequivalent representations has been recognized in problems related to quantization
of mixed particles~\cite{Mixing}. In these cases, the non-trivial nature of the physical vacuum (the so-called flavor vacuum), leads to
phenomenologically relevant corrections to the conventional  flavor oscillation formulas~\cite{BHV99}. The case of mixing is indeed emblematic in
the context of inequivalent representations: a simple combination of fields with different masses has a dramatic effect on the structure of the
Hilbert space and consequently the expectation values and Green's functions for the flavor fields have to be calculated by use of the flavor vacuum,
which is different from the one for the fields with definite masses. In this line of development, it has been also lately studied the possibility of a dynamical generation of fermion mixing
at physical level~\cite{DynMix} and the extension to curved backgrounds~\cite{Blasone:2015bea}.

Aforementioned developments have been basically achieved only in the context of canonical quantization, where there is a clear distinction between the level of dynamics (i.e., the operatorial Heisenberg equations) and that of the representation which embodies the boundary conditions for the Heisenberg equations
and defines the Hilbert space. On the other hand, one would intuitively expect that analogous results should be obtainable through the use of the functional-integral
(FI) techniques, where the emphasis is shifted from operators to partition functions, correlation functions and time-evolution kernels.

The aim of the present paper is to investigate the r\^{o}le of inequivalent representations in the functional-integral framework. In particular, we ask ourselves following questions:
\emph{Does the conventional functional integral know about inequivalent representations? If yes, how are the representations manifested in functional integrals? If not, how this can be rectified?}


So far these kinds of questions have been pursued in the literature only indirectly and in very specific contexts.
For instance, in Ref.~\cite{MatPapUme} the authors showed how the phenomenon of spontaneous symmetry breaking can be treated with the help of FIs by introducing the so-called $\epsilon$-term prescription (not to be mistaken with the
Feynman--Stuckelberg $\epsilon$ prescription). This consists of adding an explicit symmetry breaking term of the form $\mathcal{L}_\epsilon = i  \epsilon (\phi-v)^2$ to the Lagrangian, where \cite{MatPapUme} $\phi$ represents a Goldstone-like scalar (singlet) field and $v$ is a real number.  The
limit $\epsilon \rightarrow 0_+$ (for any $\epsilon > 0$) has to be taken at the end of calculations.
In this setting, it was possible to show that different values of $v$ label unitarily inequivalent Fock spaces~\cite{MatPapUme}.

Another pertinent example of inequivalent representations within FIs was presented in Ref.~\cite{Torre}. There it was studied the solution of the coherent-state evolution kernel for quadratic Hamiltonians of the form $H =  \sum_{j,k} \!\lf(A_{j \, k} z^*_j (t) z_k (t) \ri.$          $ \lf.+ \ha B_{j \, k} z_j (t) z_k (t)+\ha B^*_{j \, k} z^*_j (t) z^*_k (t)  \ri)$.  When the range of the
summation was considered to be infinite or continuous (i.e., integration) then the evolution kernel was represented by means of the coherent-state  FI. Within this setting,
it was found that a sufficient condition for the existence of the solution (i.e., when the ensuing Fredholm determinant is defined) can be expressed as
$\sum_{j,k} B_{j \, k} B^*_{j \, k} < \infty$. This is tantamount to the requirement that both the final and initial-time Fock spaces belong to the same domain of definition of the Hamiltonian operator.
Or, in other words, viewing the time evolution as the symplectic transformation generated by the above Hamiltonian, the two
canonically transformed Fock spaces are unitarily inequivalent when the above sufficiency condition is not satisfied.

More recently, it was shown in Ref.~\cite{TelNog} that some ambiguities arising in the evaluation of the partition function in QFT should be related to the existence of inequivalent representations of canonical (anti)-commutation relations. In~\cite{TelNog}, the BCS model of superconductivity was used as an illustrative example. By following a similar lines of reasoning as in Refs.~\cite{UTK,kinks}, the authors demonstrated that the choice of a bilinear part of the action functional defines the appropriate asymptotic vacuum state and hence fixes the  physical representation. In the BCS case, this choice leads to the energy-gap equation.

Our specific intent here is to demonstrate how inequivalent representations of the CCR emerge when the FIs are employed in the study of {\em linear canonical transformations}. We use linear canonical transformations for two principal reasons; (a) they represent a paradigmatic mechanism for generation of inequivalent representations in QFT~\cite{UMZ2,BJV}, (b) according to a Groenewold--van Hove theorem~\cite{Groe,VH1951} they constitute the most general class of symplectic transformations that are unitarily implementable in QM. In dealing with canonical transformations, we follow loosely the approach proposed in Ref.~\cite{Swa}, but instead of the phase-space FIs we conduct our treatment in the framework of the coherent-states FIs which are both mathematically and conceptually less problematic than their phase-space counterparts. In addition, they are also more appropriate from the QFT point of view since there the Fock space is defined by the action of annihilation and creation operators on the vacuum state~\cite{BJV,Umezawa:1982nv}
and so, the most natural framework to study the QFT in the FI setting is the one which deals with ``classical analogues'' of ladder operators~\cite{FadSlav}.

We shall see that a canonical transformation always implies a change of basis on which these operators are evaluated, and for systems with infinite number of degrees of freedom, such a change of Hilbert space might, in principle, result in the new Hilbert space that is unitarily inequivalent to the original one.
In view of this, a central object in our analysis is represented by the matrix elements of the generator of a given canonical transformation, which can be regarded as the transition amplitude among two different representations of CCR, vanishing in the case when the respective Hilbert spaces become orthogonal in the large volume limit. Such  matrix elements are often easily accessible within the operatorial formalism, but are usually not considered in the FI framework. Apart from its formal interest, our result can be used for the study of (non-unitary) transitions among inequivalent representations.

%
%
%

Here we present a self-contained exposition, where all discussed statements concerning inequivalent representations are proved by means of a PI or FI
argumentation and are illustrated by several examples. Beside new results, we also collect and generalize some older ones, which are scattered
in an incoherent fashion over a series of articles.
Furthermore, we also wish to promote the concept of inequivalent representations in QFT
which is not yet sufficiently well known among the path-integral practitioners~\cite{BJSPrague}.

The structure of the paper is as follows. To set the stage we elucidate in the next section the appearance of inequivalent representations
in the framework of canonical quantization with two simple examples: the van Hove model~\cite{vanHove} and the Bogoliubov model of a weakly interacting Bose gas~\cite{Bog}.
In Section~\ref{SEc3}, we give a brief review of known results on canonical transformations in the context of phase-space path integrals.
%
%
Since PIs represent a first-quantized version
of FIs, they serve as an important testbed for more sophisticated QFT considerations.
In Section~\ref{SEc4}, we revise the issue of canonical transformations in the framework of coherent-state PIs. For simplicity's sake, we limit our discussion to one-mode systems.
We then use the obtained PI framework to re-analyze both the van Hove model and the Bogoliubov model. This is done in Section~\ref{excan}.
In Section~\ref{SEc7}, we pass to QFT and generalize our PI findings to the field-theoretical FIs with a particular emphasis on the emergence of inequivalent representations and connection with continuous phase transitions. Finally, Section~\ref{SEc8} summarizes our results and discusses possible
extensions of the present work.
For the reader's convenience, the paper is supplemented with three appendices which clarify some finer technical details needed in the main text.

\section{Canonical transformations in QFT and inequivalent representations}\label{SEc2}

In this section we shall clarify how inequivalent representations of CCR appear in QFT  when canonical transformations are used
within the conventional operatorial formalism. To simplify our exposition, we will deal with two simple but, in a sense, characteristic examples: the van Hove model, and the weakly interacting Bose gas model developed by Bogoliubov. In Section~\ref{SEc3} we will re-analyze these examples with FI techniques.

Canonical transformations in QM have been systematically studied in a number of works (see, e.g.,  Ref.~\cite{And} and citation therein).
However, the key result was spelled out by Groenewold already in his 1946 paper~\cite{Groe} and further
refined by van~Hove in a classic work~\cite{VH1951}.
Ensuing theorem, the so-called Groenewold--van~Hove ``no-go'' theorem, states that there exists a one-to-one correspondence between classical symplectic transformations and unitary transformations of quantum theory only when the generating function is at most quadratic,
i.e., in the case of linear canonical transformations. Only in such a case, one can derive a close relationship between the quantum generator
and the generating function of the corresponding classical transformation. When the quantum generator does not exist or is infinite --- as it may happen in the QFT setting, then the classical generating functional does not exist either. This is typically reflected by various divergences obtained in
the infinite-volume limit. To regulate the computations at intermediate stages,
our subsequent considerations will be done in the so called \emph{box regularization}, i.e. inspected systems will be treated as being confined in a finite box of volume $V$.
The regulator will be removed (i.e., the large-$V$ limit will be performed)
through the identifications~\cite{BJV, Miransky}:
\bea
\sqrt{V} a_{\G k} & \rightarrow & a(\G k) (2 \pi)^{{3}/{2}}\, , \label{ptvo1} \\
\frac{1}{V}\sum_{\G k} & \rightarrow & \frac{1}{(2 \pi)^3} \intk \, , \label{ptvo2} \\
\frac{V \delta_{\G k \G p}}{(2 \pi)^3} & \rightarrow & \delta(\G k-\G p) \label{ptvo3} \, ,
\eea
only at the end of our calculations.

\subsection{van Hove model} \label{vhop}

Our first system of interest is the van Hove model which is described by the Hamiltonian
\be
\hat{H}_{vH} \ = \ \sum_\G k  \lf[\om_{\G k} \ \! \hat{a}^\dagger_{\G k} \hat{a}_{\G k} \ + \ \frac{\nu_\G k}{\sqrt{V}} \lf(\hat{a}^\dagger_{\G k} \ + \ \hat{a}_{\G k}\ri) \ri] \, .
\label{VHHam}
\ee
We define $|0\ran$ as the state annihilated by $\hat{a}_{\G k}$ for each $\G k$ via the auxiliary condition:
\be \label{vacdef}
\hat{a}_\G k |0\ran \ = \ 0 \, .
\ee
This can be regarded as the vacuum state for the free part of the Hamiltonian\footnote{This mimics precisely the way how the $in$- or $out$-state vacuum is defined in QFT via a bilinear part of the Hamiltonian, provided one can assume a cluster property of the system~\cite{Haag}.} Eq.\eqref{VHHam}. Hamiltonian Eq.(\ref{VHHam}) can be diagonalized via linear transformation (also known as the dynamical or Haag's map~\cite{BJV})
\bea
\hat{\al}_\G k &=& \hat{a}_\G k \ + \ \frac{g_\G k}{\sqrt{V}}  \, , \label{HaagmapvH}\\
\hat{\al}^\dagger_\G k &=& \hat{a}^\dagger_\G k \ + \ \frac{g_\G k}{\sqrt{V}} \, ,
\label{HaagmapvH2}
\eea
where $g_\G k = {\nu_\G k}/{\om_\G k}$ is a real function. By defining the generator of the canonical transformation
\be
\hat{G} \ = \ \exp \! \lf[\frac{1}{\sqrt{V}}\sum_{\G k} \lf(g_\G k \hat{a}^\dagger_\G k-g_\G k \hat{a}_\G k\ri)\ri] \, , \label{generator}
\ee
we can alternatively write Eqs.(\ref{HaagmapvH}),(\ref{HaagmapvH2}) as
\bea
\hat{\al}_{\G k} &=& \hat{G}^\dagger \, \hat{a}_{\G k} \, \hat{G} \, ,  \label{neope1}\\[1mm]
\hat{\al}^\dagger_{\G k} &=& \hat{G}^\dagger \, \hat{a}^\dagger_{\G k} \, \hat{G}   \label{neope2} \, .
\eea
Note that the operator $\hat{G}$ is, by its very form, unitary for any finite $V$. With Eqs.(\ref{HaagmapvH})-(\ref{HaagmapvH2})
[or equivalently Eqs.(\ref{neope1}),(\ref{neope2})] we can cast $\hat{H}_{vH}$ in the diagonal form
\begin{eqnarray}
\hat{H}_{vH} \ = \ \sum_\G k \om_{\G k}\!\left(\hat{\al}^\dagger_\G k \hat{\al}_\G k \ - \ \frac{g_\G k^2}{V}\right).
\end{eqnarray}
The associated {\em physical} vacuum $|0(g) \ran$ is defined by the auxiliary condition
\be
\hat{\al}_\G k |0(g) \ran \ = \ 0 \, .
\ee
This is related with  $|0\ran$ through the relation
\be
|0(g)\ran \ = \ \hat{G}^{-1}|0\ran \, .
\ee
The overlap between the two vacua is thus
\be
\lan 0 | 0(g) \ran \ = \ \lan 0 | \hat{G}^{-1}|0\ran \ = \ \exp\lf(-\frac{1}{2 V } \sum_k g_\G k^2\ri).
\ee
In the large-$V$ limit, we can employ prescriptions Eqs.\eqref{ptvo1}-\eqref{ptvo3} to obtain
\be  \label{vvvH}
\lan 0 | 0(g) \ran \ = \ \exp\lf[-\ha\intk \ \! (g (\G k))^2\ri].
\ee
Here we have defined $g (\G k)=g_\G k (2 \pi)^{3/2}$. Let us now consider the important special case of  translational invariance, i.e. $\nu (\G k)=\nu \,\delta(\G k)$ and then $g(\G k)=c \, \delta(\G k)$. Therefore the scalar product Eq.(\ref{vvvH}) can be explicitly written as
\be
\lan 0 | 0(g) \ran=\exp\lf(-\ha \frac{V c^2}{(2 \pi)^3}\ri).
\ee
In the thermodynamical limit this goes to zero, i.e. the Hilbert spaces constructed over the respective vacuum states are orthogonal. From the second Schur's lemma~\cite{Tung} it then follows that the two representations of CCR (Weyl-Heisenberg algebra) cannot be connected by a unitary transformation.

Let us note, that the assumption of  translational invariance used in the derivation of the unitary inequivalence was helpful but not necessary. Indeed, any $g (\G k)$ that is not a square-integrable function
will lead to the same conclusion. On the other hand, the canonical transformation Eqs.(\ref{HaagmapvH})-(\ref{HaagmapvH2}) with  $g (\G k) \in L^2({\G {R}}^{\!3})$ does not lead to two unitarily inequivalent physical systems.

\subsection{Bogoliubov model}

Our second system of interest is the model for a weakly interacting Bose gas originally proposed by Bogoliubov~\cite{Bog,Miransky,LiPi}.
The ensuing Hamiltonian reads
\bea \label{Bogham}
\mbox{\hspace{-2mm}}\hat{H}_B \ = \ \sum_{\G k} \frac{k^2}{2m}\hat{a}^\dagger_\G k \hat{a}_\G k \ + \ \frac{N^2}{2V}U_0
 \ + \ \frac{N}{2V}U_0\sum_{\G k \neq \G 0}\lf[\hat{a}^\dagger_\G k \hat{a}^\dagger_{-\G{k}}+\hat{a}_\G k \hat{a}_{-\G{k}}+2 \hat{a}^\dagger_\G k \hat{a}_\G k \ri] .
\eea
Here $N$ denotes the number of $\G k$ modes. $\hat{H}_B$ can be diagonalized by a \emph{Bogoliubov}--\emph{Valatin transformation}:
\begin{eqnarray}
&&\hat{\al}_\G k \ = \ u_\G k \, \hat{a}_\G k \ + \ v_\G k \, \hat{a}^\dagger_{-\G k}\, , \label{dynmapsupa} \\
&&\hat{\al}^\dagger_\G k\ = \ u_\G k \, \hat{a}^\dagger_\G k \ + \ v_\G k \, \hat{a}_{-\G k}\, , \label{dynmapsup}
\end{eqnarray}
with
\be
u^2_\G k \ - \ v^2_\G k \ = \ 1 \, .
\ee
In order to diagonalize the Hamiltonian Eq.\eqref{Bogham}, the parameters $u_\G k$ must take the form
\begin{subequations}
\be
u_\G k \ = \ \frac{1}{\sqrt{1-L^2_\G k}}\, , \label{uka}
\ee
\be
v_\G k\  = \frac{L_\G k}{\sqrt{1-L^2_\G k}}\, , \label{ukb}
\ee
\end{subequations}
where
\be
L_\G k\ = \ \frac{1}{mu^2}\lf[\frac{{\G k}^2}{2m} \ + \ mu^2\ - \ \ep({\G k})\ri], \label{enspequa1}
\ee
and
\bea
\ep({\G k})&=&\sqrt{u^2 {\G k}^2 \ + \ \lf(\frac{{\G k}^2}{2m}\ri)^2} \, ,\label{enspequa} \\
u&=&\sqrt{\frac{U_0 N}{m V}}\, .
\eea
Consequently, the transformed Hamiltonian takes the form
\be
\hat{H}_B \ = \ \sum_{\G k} \ep({\G k}) \hat{\al}^\dagger_\G k \hat{\al}_\G k \ + \ E_0\, , \label{phyboham}
\ee
with
\be \label{zeropoint}
E_0\ = \ \ha N m u^2 \ + \  \ha \sum_{\G k\neq \G 0} \lf[\ep({\G k})\ - \ \frac{{\G k}^2}{2m}-mu^2 \ + \ \frac{m^3 u^4}{{\G k}^2}\ri].
\ee
Therefore Eq.\eqref{dynmapsup} represents the dynamical map of the system: here $ \hat{\al}^\dagger_\G k$ and $ \hat{\al}_\G k$ are the creation and annihilation operator, respectively, for the physical excitations (quasi--particles) of the system. Their energetic spectrum is given by Eq.\eqref{enspequa}.

The vacua $|0\ran$ and $|0(\theta)\ran$ are connected by the relation
\be \label{bogvac}
|0(\theta)\ran \ = \ \hat{G}^{-1}|0\ran\, ,
\ee
where $\hat{G}$ is the generator of the Bogoliubov--Valatin transformation Eq.\eqref{dynmapsup}
\be
\hat{G} \ = \ \exp \lf[\sum_\G k \theta_\G k \lf( \hat{a}^\dagger_\G k \hat{a}^\dagger_{-\G k} \ - \ \hat{a}_\G k \hat{a}_{-\G k} \ri) \ri] ,
\ee
and
\be
u_\G k \ = \ \cosh \theta_\G k \, , \,\,\,\,\,\,\,\,\, v_\G k \ = \ \sinh \theta_\G k \, .
\ee
With the help of the Gaussian decomposition~\cite{Perelomov}, we can rewrite the relation~\eqref{bogvac} in the large-$V$ limit as
\bea
|0(\theta)\ran = \exp \lf[-\frac{V}{(2 \pi)^3} \intk \ \log \cosh \theta(\G k)\ri]
 \exp \lf[ \intk \ \tanh \theta(\G k) \hat{a}^\dagger_\G k \hat{a}^\dagger_{-\G k}\ri] |0\ran  \, .
\eea
Consequently, the overlap between the two vacua is
\be \label{vvB}
\lan 0|0(\theta)\ran \ = \ \exp \lf[-\frac{V}{(2 \pi)^3} \intk \ \log \cosh \theta (\G k)\ri] ,
\ee
which clearly goes to zero as $V$ goes to infinity, in agreement with the Haag theorem \cite{Haag,StWi}.
The two representations of CCR are thus unitarily inequivalent in the field-theory limit.
Let us stress finally, that in contrast to the van Hove model, here the unitary inequivalence holds true for any function $\theta (\G k)$
(apart from the trivial case $\theta (\G k)=0$).

\section{Canonical transformations in phase-space path integral}\label{SEc3}

In this section we briefly review the theory of canonical transformations in the framework of phase-space PIs. This is a particularly instructive starting point for our subsequent discussion of coherent-state PIs.
Due to an extensive literature related to phase-space PIs  (see, e.g., Refs.~\cite{Fan,GeJe,Schu1982,FuKa,Shi}
and citation therein) one can often easily foresee a number of
subtleties that can be anticipated in FIs in connection with canonical transformations.
Our exposition here follows closely the approach to canonical transformations developed in Ref.~\cite{Swa}.
There the key object of interest is a mixed (the so-called Kirkwood) evolution kernel (here and throughout $\hbar = 1$)
\be
W_{fi}\ = \ \lan p_f |\exp \lf[-i \hat{H}(\hat{p},\hat{q};t)(t_f-t_i)\ri] |q_i \ran \, .
\ee
This can be represented in the time-sliced PI form~\cite{Swa,Kle}
\begin{eqnarray}
W_{fi} \ =  \ \lim_{N\rightarrow \infty}\mbox{\hspace{-6mm}} &&\frac{1}{\sqrt 2 \pi} \prod^N_{j=1} \int \frac{\dr p_j}{2 \pi} \dr q_j \exp\left(-i q_i p_1 \right)\nonumber \\
&&\times \ \exp \lf\{-i \sum^N_{j=1}\lf[ q_j (p_{j+1}-p_j)+ H(p_j,q_j) \Delta t\ri] \ri\} \nonumber \\
\ =  \ \lim_{N\rightarrow \infty}\mbox{\hspace{-6mm}} &&\frac{1}{\sqrt 2 \pi} \prod^N_{j=1} \int \frac{\dr p_j}{2 \pi} \dr q_j \exp\left(-i q_N p_f \right)\nonumber \\
&&\times \ \exp \lf\{i \sum^N_{j=1}\lf[ p_j (q_{j}-q_{j-1}) -  H(p_j,q_j) \Delta t\ri] \ri\} ,
\label{pspath}
\end{eqnarray}
where $p_{N+1}\equiv p_f$ and $q_0 \equiv q_i $ is understood.

To define a quantum mechanical canonical transformation, one starts from the (Schr\"{o}dinger-picture) resolution of unity
\be
\ide \ = \ \int \frac{\dr P_m}{\sqrt{2\pi}} \dr Q_n  \ \!|Q_n \ran e^{i P_m Q_n}\lan P_m | \, ,
\label{compswb}
\ee
which can then be employed to write
\be
\lan p_f |q_i \ran \ = \ \int \frac{\dr P_m}{\sqrt{2\pi}} \dr Q_n \ \! \lan p_f |Q_n \ran e^{i P_m Q_n}\lan P_m |q_i \ran \, .
\label{compswa}
\ee
One can make the following ansatz for the mixed-state products
\bea
\lan p_f |Q_n \ran &=&\frac{1}{\sqrt{2 \pi}}\exp\lf\{i\lf[P_f(Q_f-Q_n)+F(p_f,Q_f)\ri]\ri\} \, , \label{ansswa1} \\
\lan P_m |q_i  \ran &=&\frac{1}{\sqrt{2 \pi}}\exp\lf\{-i\lf[P_m Q_i+F(p_i,Q_i)\ri]\ri\} \, ,  \label{ansswa2}
\eea
where $F(p,Q)$ is some function to be specified. To find $F(p,Q)$, we substitute Eqs.(\ref{ansswa1})-(\ref{ansswa2}) in Eq.\eqref{compswa}. This yields
\be
\lan p_f |q_i \ran \ = \ \frac{1}{\sqrt{2 \pi}} \exp \lf\{i\lf[P_f(Q_f-Q_i)+F(p_f,Q_f)-F(p_i,Q_i)\ri]\ri\} .
\ee
To preserve the consistency of this equation, one has to impose
\bea
P_f \lf(Q_f-Q_i\ri)&=& F(p_f,Q_i) - F(p_f,Q_f) \, ,  \label{postswa1} \\
q_i \lf(p_f-p_i\ri)&=& F(p_i,Q_i) - F(p_f,Q_i)  \, .    \label{postswa2}
\eea
In this way we obtain
\be
\lan p_f |q_i \ran \ = \  \frac{1}{\sqrt{2 \pi}} \exp[-i q_i(p_f-p_i)]\, .
\ee
Clearly, this is the correct  only when $p_i=0$ and/or $q_i=0$. However, one cannot impose the condition $p_i=0$ because this would
over-specify the  problem and it would be also at odds with the Heisenberg uncertainty principle which requires
that we cannot know both $q_i$ and $p_i$.
In the case when $q_i\neq 0$ one can still use the former prescription. As discussed in Ref.~\cite{Swa}, this can be most easily done in the framework  of the WKB approach where the non-trivial boundary conditions are infused into a phase factor and the fluctuation factor
is now a PI with boundary conditions $\delta q_i=0$ and $\delta p_f=0$ which are clearly in accord with the above consistency relation and can be thus treated with the previous technique for canonical transformations.

Relations Eqs.\eqref{postswa1},\eqref{postswa2} define a new couple of phase-space variables $Q$ and $P$.
To see this, we observe that Eqs.\eqref{postswa1},\eqref{postswa2} hold for any initial and final times and thus also for two adjacent time slices. Then
\bea
P_j &=& -\frac{F(p_j,Q_j)-F(p_j,Q_{j-1})}{\Delta Q_j} \, , \label{postswa3} \\
q_j &=&-\frac{F(p_{j+1},Q_j)-F(p_j,Q_j)}{\Delta p_j} \, ,    \label{postswa4}
\eea
where $\Delta Q_j=Q_j-Q_{j-1}$ and $\Delta p_j=p_{j+1}-p_j$.
%
In addition, from Eqs.\eqref{postswa3},\eqref{postswa4}, it follows that
\be
q_j (p_{j+1}-p_j) \ = \ F(p_j,Q_j) \ - \ F(p_{j+1},Q_{j+1}) \ - \ P_{j+1}(Q_{j+1} -  Q_j) \, ,
\ee
and hence the argument of the PI Eq.\eqref{pspath} can be rewritten as
\bea
&&\mbox{\hspace{-15mm}} -\sum^N_{j=0}\lf[q_j(p_{j+1}-p_j)\ + \ \epsilon H(q_j,p_j)\ri] \ = \  F(p_f,Q_f) \ - \ F(p_i,Q_i)\nonumber \\
&&\mbox{\hspace{15mm}}+ \ \sum^N_{j=0}\lf[P_{j+1}(Q_{j+1}-Q_j)-\epsilon H(P_j,Q_j,\Delta P_j,\Delta Q_j)\ri].
\label{postswa4b}
\eea
Eqs.\eqref{postswa3},\eqref{postswa4} and Eq.\eqref{postswa4b} indicate that $F$ might be identified with a time-sliced version of the type-3 generating function for canonical
transformations \cite{Goldstein}. By Taylor expanding the RHSs of Eqs.\eqref{postswa3},\eqref{postswa4} we get
\begin{eqnarray}
P_j &=&  -\frac{\pa F(p_j,Q_j)}{\pa Q_j}\ + \ \ha \frac{\pa^2 F(p_j,Q_j)}{\pa Q^2_j}\ \!\Delta Q_j\ + \ O(\Delta Q_j^2)\, ,\nonumber \\
q_j &=& -\frac{\pa F(p_j,Q_j)}{\pa p_j}\ - \ \ha \frac{\pa^2 F(p_j,Q_j)}{\pa p^2_j}\ \! \Delta p_j \ + \ O(\Delta p_j^2)\, .
\label{tasv}
\end{eqnarray}
So, in the limit $\epsilon = \Delta t  \rightarrow 0$  (or equivalently $N \rightarrow \infty$) this allows to identify $F$ with the type-$3$ (classical) generating function of canonical transformations as the leading-order behavior coincides with the ensuing classical Legendre-transform relations. However\footnote{In Ref.~\cite{Schu1982} it was shown that, if we extend our considerations to time-dependent transformations and choose $F(p,Q)$ satisfying the classical Hamilton--Jacobi equation, we would always obtain the semiclassical solution.}, a careful analysis of a full time-sliced representations of PIs reveals that there are two complications related with the above interpretation; (a) the time-sliced canonical transformation generates in the action additional terms that are of order $O(\Delta Q_j)$ i.e., terms that need not vanish in the continuous limit  $\epsilon \rightarrow 0$ --- the so-called Edwards--Gulyaev anomaly~\cite{EG}, (b) the PI phase-space measure cannot be viewed as a product of Liouville measures and, as a rule, canonical transformations often produce the so-called Liouville anomaly~\cite{Swa} --- the Jacobian is not unity. In fact, it is not difficult to see, using Eqs.\eqref{postswa1},\eqref{postswa2} that in the case of quantum transformations, in the inverse Jacobian
\be
J^{-1} \ = \ \prod^N_{j=1} \lf(\frac{\pa Q_j}{\pa q_j}\frac{\pa P_j}{\pa p_j}-\frac{\pa P_j}{\pa q_j}\frac{\pa Q_j}{\pa p_j}\ri) \, ,
\ee
appear the so-called \emph{anomalous (Liouville) corrections}~\cite{Swa,GeJe,FuKa,Shi,SchuMc}, namely
\be \label{jacps}
J^{-1}\ = \ \prod^N_{j=1}\lf[1+A_J\Delta Q_j+B_J\Delta P_j\ri] \, ,
\ee
where, following Ref.~\cite{Swa}, we have introduced the shorthand notation
\bea
A_j&=& \ha\lf(\frac{\pa^3 F}{\pa p^2_j \pa Q_j}\frac{\pa Q_j}{\pa q_j}\frac{\pa p_j}{\pa Q_j}+\frac{\pa^3 F}{\pa Q^2_j \pa p_j}\frac{\pa Q_j}{\pa q_j}\ri) \, , \\
B_j&=& \ha \frac{\pa^3 F}{\pa p^2_j \pa Q_j}\frac{\pa Q_j}{\pa q_j}\frac{\pa p_j}{\pa P_j} \, .
\eea
When $\Delta Q_j=O(\Delta t)$ and/or $\Delta P_j=O(\Delta t) $ these terms give a finite contribution. In fact, we can recast the expression Eq.\eqref{jacps} in the form
\begin{eqnarray}
J^{-1} &=& \exp\!\left[\log \prod^{N}_{j=1} (1+A_j \Delta Q_j+B_j\Delta P_j)\right] \nonumber \\
&\sim & \exp\!\left[\sum^{N}_{j=1}(A_j \Delta Q_j+B_j\Delta P_j)\right] ,
\label{Jacexph}
\end{eqnarray}
where, on the second line we have expanded the logarithm to the first order in increments of $P$ and $Q$. Note that Eq.\eqref{Jacexph} can be equivalently rewritten as
\be
J^{-1}\ = \ \exp\left[{\sum^{N}_{j=1}\Delta t \lf(A_j \frac{\Delta Q_j}{\Delta t}+B_j\frac{\Delta P_j}{\Delta t}\ri)}\right] ,
\ee
which in the large $N$ (or long-wave) limit reduces to
\be
J^{-1}\ = \ \exp\left[{\int^{t_f}_{t_i} \dr t \lf(A(t) \dot{Q}(t)+B(t) \dot{P}(t)\ri)} \right]  .
\label{IV.55a}
\ee
When $J \neq 1$ one gets an extra contribution from Eq.(\ref{IV.55a}) in the PI action. In this case one speaks about the {\em Liouville anomaly}. For coherent-state PIs a similar situation will be discussed in Appendix~C.

\section{Canonical transformations and coherent-state path integral} \label{SEc4}

In this section we extend our previous discussion to the study of canonical transformations in the framework of coherent-state PIs. The latter are also known as PIs in holomorphic representation~\cite{ItzZub,FadSlav}.
Apart from the issue of canonical transformations, our focus on the coherent-state PIs is dictated by some other (mostly conceptual) issues. First of all, as mentioned in the Introduction, we are interested in
properties that are related with the structure of Hilbert space emerging from various (inequivalent) representations of the Weyl--Heisenberg algebra (i.e., algebra of annihilation and creation operators). In this respect, it is natural to use a framework in which the ladder operators are diagonal from very scratch.

Second, as pointed out by a number of authors~\cite{Schweber,KlaSka,Klauder},  coherent-state PIs can be expressed in terms of a formal Gaussian measure which makes them less pathological than their phase-space counterparts. Moreover, this measure is naturally symmetric in the integration variables\footnote{For instance, in the phase-space PIs for $q$-$q$ transition amplitudes there is (in the times-sliced form) one more integration over $p$ than over $q$.}, thus avoiding problems associated with the asymmetry of the phase-space PI measure and boundary conditions, which one meets when trying to define canonical transformations~\cite{Schu1982} (see Section~\ref{SEc3}). This advantage was already recognized in Ref.~\cite{Swabook}.

We  show that the difficulties enumerated above do not occur when dealing with coherent-state PIs. This framework will facilitate the discussion of systems with infinite number of degrees of freedom, where inequivalent representations are known to occur. Actually, QFT represents a prototype context where FIs in holomorphic representation are widely used~\cite{ItzZub,Kle,FadSlav}.

The basic propositions of this section are formulated in terms of single-mode quantum systems, which make the ensuing discussion and conclusions shorter and more apparent. In addition, the general nature of the results obtained will be crucial in Sections~\ref{excan} and~\ref{SEc7}.
For the reasons mentioned in the Introduction we will limit our consideration here only to the linear canonical transformations.

To give some necessary details on the holomorphic representation of PIs we start by reminding that a Glauber (or Weyl--Heisenberg) coherent state $|z\ran$  is defined as~\cite{Perelomov}
\be
|z\ran \ = \ e^{z \hat{a}^\dagger}|0\ran \, , \qquad
\lan z^*| \ = \ \lan 0|e^{z^* \hat{a}} \, .
\ee
Here we use the definition of coherent states which are not normalized~\cite{BJV}. In this way many subsequent expressions get simpler form.
With this we have
\bea
\hat{a}  |z\ran \  = \ z |z\ran \, , \qquad \lan z^*| \hat{a}^\dag \ = \ z^*\lan z^*| \, .
\eea
The resolution of unity takes the form
\be
\ide \ = \ \int \dr \mu(z) |z\ran \lan z^*| \, ,
\ee
where the integration measure is defined as
\be
\dr \mu(z)  \ \equiv \ \frac{\dr z \dr z^*}{2\pi i}e^{-z^* z}\, .
\ee

Let us now consider a system with Hamiltonian $\hat{H}(\hat{a},\hat{a}^\dagger)$, written in terms of annihilation and creation operators. For simplicity's sake we do not consider here Hamiltonians with an explicit time dependence. Defining the \emph{moving (Heisenberg picture) base vectors}
\be
|z,t\ran \ = \ e^{i \hat{H} t}|z\ran \, ,
\ee
the resolution of the identity can be cast as
\be
\int \dr \mu(z) \ \! |z^*,t\ran\lan z,t|\ = \ \ide \, .
\ee
Following the same route as in the usual $q$ configuration-space, the evolution kernel $\lan z^*_f,t_f|z_i,t_i\ran$ can be written in the PI representation as~\cite{BJV}
\be \label{initpathint}
\hspace{-0.5 cm}\lan z^*_f,t_f|z_i,t_i\ran \ = \ \int^{z^*(t_f)=z^*_f}_{z(t_i)=z_i} \mathcal{D}z^* \mathcal{D}z \ \! e^{z^*_f z_f}e^{i\!\int^{t_f}_{t_i} \dr t [i z^*(t) \dot{z}(t)-H(z^*,z)]} \, .
\ee
This PI representation is also known as the {\em holomorphic representation}~\cite{ItzZub,FadSlav} and it is just a shorthand form for the time-sliced expression
\bea
\lan z^*_f,t_f|z_i,t_i\ran &=& \lim_{N\rightarrow \infty} \prod^{N}_{j=1} \int \frac{\dr z_j^* \dr z_j}{2 \pi i} \  \! e^{z^*_f z_f}e^{-\sum^{N+1}_{j=1}z^*_j(z_j-z_{j-1})} \non \\
 & \times & e^{-i\sum^{N}_{j=0}H(z^*_{j+1},z_j)\Delta t}\, ,  \label{feynker}
\eea
where
\be
H(z^*_{j+1},z_j,t_j) \ \equiv \ \frac{\lan z^*_{j+1},t_{j+1}|\hat{H}(\hat{a}^\dagger,\hat{a})|z_j, t_j \ran}{\lan z^*_{j+1},t_{j+1}|z_j, t_j \ran} \, .
\ee
Let us now consider another set of coherent states $|\zeta\ran\!\ran$ spanning a Hilbert space $\tilde{\mathcal{H}}$. The latter is constructed over a new vacuum state $|0 \ran\!\ran$ by an action of new ladder operators $\hat{\al}$ and $\hat{\al}^\dagger$ defined by\footnote{We use the notation $|0\ran \!\ran$ for consistency with Section \ref{SEc7} where the QFT framework is considered.}
\bea
\hat{\al} \ = \ \hat{G}^\dagger \, \hat{a} \, \hat{G} \, ,  \label{neope3} \qquad \hat{\al}^\dagger \ = \ \hat{G}^\dagger \, \hat{a}^\dagger \, \hat{G}   \, ,
\eea
as
\bea
|\zeta \ran\!\ran = e^{\zeta \hat{\al}^\dagger}|0 \ran\!\ran\, , \qquad \lan\!\lan \zeta^*| = \lan\!\lan 0|e^{\zeta^* \hat{\al}} \, .
\eea
These are eigenstates of new annihilation and creation operators
\be
\hat{\al}  |\zeta\ran\!\ran \ = \ \zeta |\zeta\ran\!\ran \, ,\qquad \,\, \lan\!\lan \zeta^*| \hat{\al}^\dag \ = \ \zeta^*\lan\!\lan \zeta^*| \, .
\ee
We can write
\be
\hat{G} \ = \ \exp\lf[-i \theta \hat{K}\lf(\hat{a}^\dagger,\hat{a}\ri)\ri] \, ,
\label{66a}
\ee
where $\te$ is real and $\hat{K}$ is self-adjoint. As seen in Section~\ref{SEc3}, the starting point for the discussion of canonical transformations are mixed scalar products
\be
\lan\!\lan \zeta^* | z \ran \ = \ \sum^\infty_{n=0}\frac{\zeta^{* \, n} }{n!}\ \! \lan\!\lan 0 |\hat{\al}^{n}|z \ran\, .
\ee
By using Eq.\eqref{neope3}, this can be written in a more explicit form as
\be
\lan\!\lan \zeta^* | z \ran  \ = \ \sum^\infty_{n=0}\frac{\zeta^{* \, n} }{n!}\lan\!\lan 0 | \lf(\hat{G}^\dagger \hat{a} \hat{G}\ri)^n|z \ran  \, .
\ee
Moreover, using the fact that
\be
\lf(\hat{G}^\dagger \hat{a} \hat{G}\ri)^n \ = \ \hat{G}^\dagger \hat{a}^{n} \hat{G} \, ,
\ee
we have
\be
\lan\!\lan \zeta^* | z \ran \ = \ \sum^\infty_{n=0}\frac{\zeta^{* \, n} }{n!}\lan 0 |\hat{a}^{n} \hat{G} |z\ran\, .
\ee
Actually, this is nothing but a matrix element of the quantum generator, i.e.
\be \label{genex}
\lan\!\lan \zeta^* | z \ran \ = \ \lan \zeta^* | \hat{G} | z \ran \, ,
\ee
where now
\be
\lan \zeta^*|\ = \ \lan 0|\exp\lf(\zeta^* \hat{a}\ri) \, .
\ee
An expression given Eq.\eqref{genex} can be rewritten in the PI form as
\be \label{Intgen}
\lan z^*|\hat{G}|\zeta\ran \ = \ \int_{\xi(0)=\zeta}^{\xi^*(\theta)=z^*} \mathcal{D} \xi \mathcal{D} \xi^* e^{\xi^*(\theta)\xi(\theta)} \, e^{i\int^\theta_0 \dr \theta' \lf[i\xi^* \frac{\dr \xi}{\dr \theta'}-K(\xi^*,\xi)\ri]} \, .
\ee
Where $K(\xi^*,\xi)$ plays the r\^{o}le of Hamiltonian.
Following Refs.~\cite{Fan,Fan1975} we seek for a solution in the form
\be \label{formsearch}
\lan \zeta^*|\hat{G}|z\ran\ = \ A (\zeta^*,z)\exp\lf[ F(\zeta^*,z)\ri] \, ,
\ee
where $A(\zeta^*,z)$ and $ F(\zeta^*,z)$ are some as yet undetermined complex functions. By solving the integral Eq.\eqref{Intgen} in the saddle-point (i.e., WKB) approximation we obtain in the present case (linear canonical transformations) an exact result. To this end, we introduce an action-like functional
\be
W\lf(\zeta^*, z \ri)\ = \ \int^{\theta}_0 \dr \theta' \lf[i \xi^*(\theta') \frac{\dr \xi(\theta')}{\dr \theta'}-K[\xi(\theta'),\xi^*(\theta')]\ri] \, .
\ee
It can be checked that
\be
W\lf(\zeta^*, z \ri) \ = \ F_2\lf(\zeta^*, z \ri) \, ,
\ee
where $F_2(\zeta^*,z)$ is the classical type-$2$ generating function of the canonical transformation in the complex (classical) phase-space mechanics \cite{Strocchi}.
This is a generalization of results found in Refs.~\cite{Testa1970,Testa1973} to the case of complex classical dynamics.
The last ingredient is the formula of the semiclassical kernel \cite{BdeKKS}:
%
\be  \label{sempropagator1}
\lan \zeta^* , \theta| z , 0 \ran \ = \ \sqrt{i\frac{\pa^2 W}{\pa \zeta^* \pa z}}\ \! e^{i {W}(\xi^c,\xi^{* \, c})}\exp\lf(\frac{i}{2} \int^{\theta}_{0} \dr \theta' \frac{\pa^2 K(\xi^c,\xi^{* \, c})}{\pa \xi \pa \xi^*}\ri),
\ee
where $\xi^{c}$ and $\xi^{* \, c}$ are solutions of ``classical'' equations of the motion
\bea
\frac{\dr \xi}{\dr \theta} \ = \ -i \frac{\pa K(\xi^{*},\xi)}{\pa \xi^*}  \, , \qquad
\frac{\dr \xi^{*}}{\dr \theta}\ = \ i \frac{\pa K(\xi^{*},\xi)}{\pa \xi} \, \label{pseduhameq2} .
\eea
With this we can identify
\bea \label{staphas1}
A\lf(\zeta^*, z \ri) &=&\lf(\frac{\pa^2 F_2\lf(\zeta^*, z\ri)}{\pa z \pa \zeta^*}\ri)^\ha \, ,  \label{relsem1}\\
F\lf(\zeta^*, z \ri) &=& F_2\lf(\zeta^*, z \ri)\ + \ \frac{i}{2} \int^{\theta}_{0} \dr \theta' \frac{\pa^2 K(\xi^c,\xi^{* \, c})}{\pa \xi \pa \xi^*} \,  . \label{relsem2}
\eea
Latter two results are really close to those presented in Refs.~\cite{Wei,Hel1977}, apart from the second piece\footnote{This is not surprising because the square root term in Eq.\eqref{sempropagator1} is not a Pauli--van Vleck--Morette determinant and has to be compensated by this correction.} on the RHS of Eq.\eqref{relsem2}. Among other things, the aforementioned correction gives  the correct zero point energy contribution for the Bogoliubov model, studied in Section~\ref{excan2}.

Similarly to Eq.\eqref{formsearch}, we write:
\be
\lan z^*|\hat{G}|\zeta\ran\ = \ A^*(z^*, \zeta)\exp\lf[F^*(z^*, \zeta)\ri] \, .
\ee

Let us now introduce the function
\be \label{staphasn}
\mathcal{G}\lf(z, \zeta^*\ri)\ = \ \log A\lf(z, \zeta^*\ri)+F\lf(z, \zeta^*\ri) \, ,
\ee
which allows to recast the above mixed products to the form
\bea
\lan z^*|\zeta \ran\!\ran \ = \ \exp \lf[\mathcal{G}^*\lf(z^*, \zeta\ri)\ri] \, , \qquad
\lan\!\lan \zeta^*|z \ran \ = \ \exp \lf[\mathcal{G}\lf(\zeta^*,z\ri)\ri] \label{mixpro2}\, .
\eea

Let us start to evaluate the ratio $\lan\!\lan \zeta^*_{j+1} | z_{j+1} \ran / \lan\!\lan \zeta^*_{j+1} | z_{j} \ran $. Remember that now we are working with a basis in the Schr\"odinger picture, which is time independent. The labels $j+1$ and $j$ only identify different complex numbers.
\be
\frac{\lan\!\lan \zeta^*_{j+1} | z_{j+1} \ran}{\lan\!\lan \zeta^*_{j+1} | z_{j} \ran} \ = \ e^{\mathcal{G}(\zeta^*_{j+1},z_{j+1})-\mathcal{G}(\zeta^*_{j+1},z_{j})} \, .
\ee
Multiplying both members for $\lan z^*_{j+1} | \zeta_{j+1} \ran\!\ran \lan\!\lan \zeta^*_{j+1} | z_{j} \ran $ and integrating over $\dr \mu (\zeta_{j+1})$ we obtain
\be \label{strangerel}
e^{z^*_{j+1} z_{j+1}} \ = \ \int \dr \mu (\zeta_{j+1}) \lan z^*_{j+1} | \zeta_{j+1} \ran\!\ran \lan\!\lan \zeta^*_{j+1} | z_{j} \ran e^{\mathcal{G}(\zeta^*_{j+1},z_{j+1})-\mathcal{G}(\zeta^*_{j+1},z_{j})} \, .
\ee
In order to fulfill Eq.\eqref{strangerel} we have to assume the following consistency relation:
\be
\mathcal{G}(\zeta^*_{j+1},z_{j+1})-\mathcal{G}(\zeta^*_{j+1},z_{j})\ = \ z_{j+1}^* (z_{j+1}-z_{j}) \, . \label{dscthirdgen1}
\ee
Evaluating now $\lan\!\lan \zeta^*_{j+1} | z_{j} \ran / \lan\!\lan \zeta^*_{j} | z_{j} \ran $, one finds
\be
\frac{\lan\!\lan \zeta^*_{j+1} | z_{j} \ran}{\lan\!\lan \zeta^*_{j} | z_{j} \ran} \ = \ e^{\mathcal{G}(\zeta^*_{j+1},z_{j})-\mathcal{G}(\zeta^*_{j},z_{j})}\, .
\ee
Multiplying both members for $\lan\!\lan \zeta^*_{j} | z_{j} \ran \lan z^*_{j} | \zeta_{j} \ran\!\ran $ and integrating over $\dr \mu (z_{j})$ we obtain
\be
e^{\zeta^*_{j+1} \zeta_{j}} \ = \ \int \dr \mu (z_{j}) \lan\!\lan \zeta^*_{j} | z_{j} \ran \lan z^*_{j} | \zeta_{j} \ran\ \! e^{\mathcal{G}(\zeta^*_{j+1},z_{j})-\mathcal{G}(\zeta^*_{j},z_{j})} \, .
\ee
We are thus led to another consistency relation
\be
\mathcal{G}(\zeta^*_{j+1},z_{j}) \ - \ \mathcal{G}(\zeta^*_{j},z_{j}) \ = \ \zeta_{j}(\zeta^*_{j+1}-\zeta^*_{j}) \, . \label{dscthirdgen2}
\ee
It is important to note that these relations have been derived under the assumption that mixed products Eq.\eqref{mixpro2} are different from zero. This is justified by noticing that second derivatives of $F_2$, appearing above, is not zero for linear canonical transformations.

By using Eqs.\eqref{staphas1}-\eqref{staphasn} and the linearity of canonical transformations\footnote{As we shall discuss in detail in Section~\ref{ptqft}, in this case $\mathcal{G}=F_2+C$, where $C$ is a constant, that can be determined in quantum theory.}, we can recast above consistency relations, in the form:
\bea
 F_2(\zeta^*_j,z_j)\ - \ F_2(\zeta^*_j,z_{j-1}) & = & z_j^* (z_j-z_{j-1}) \, ,  \label{discinftrans1} \\[2mm]
 F_2(\zeta^*_j,z_{j-1})\ - \ F_2(\zeta^*_{j-1},z_{j-1}) & = & \zeta_{j-1}(\zeta^*_j-\zeta^*_{j-1}) \, . \label{discinftrans}
\eea
Actually, this is the definition of our new variables. Taylor expanding Eq.\eqref{discinftrans1} in $z_j-z_{j-1}$ and~\eqref{discinftrans} in $\zeta^*_{j+1}-\zeta^*_j$ we obtain
\bea
\label{tayleq}
z^*_j &=& \frac{\pa F_2(\zeta^*_j,z_j)}{\pa z_j} \ + \  \ha\frac{\pa^2 F_2(\zeta^*_j,z_j)}{\pa z^2_j}\ \!(z_{j-1}-z_j) \, ,  \\
\zeta_j &=&\frac{\pa F_2(\zeta^*_{j},z_j)}{\pa \zeta^*_{j}} \ + \ \ha\frac{\pa^2 F_2(\zeta^*_{j},z_j)}{\pa \zeta^{* \, 2}_{j}}\ \!(\zeta^*_{j+1}-\zeta^*_j) \, ,\label{tayleq2}
\eea
where the second relation has to be inverted to find $\zeta^*_j$. Note that the leading order term is the classical contribution. Moreover, the latter result is exact for linear canonical transformations.

From Eqs.\eqref{discinftrans1}-\eqref{discinftrans} it follows that
\be \label{discgenrel}
\sum^{N+1}_{j=1} z_j^* (z_j-z_{j-1})\ = \ F_2(\zeta^*_f,z_f)-F_2(\zeta^*_i,z_i)\ - \ \sum^{N+1}_{j=1}\zeta_{j-1}(\zeta^*_j-\zeta^*_{j-1})\, .
\ee
The argument of the time-sliced kernel Eq.\eqref{feynker}, can be thus rewritten in  terms of new variables as\footnote{For simplicity's sake we do not consider here generating functions with explicit time dependence. So, we regard only restricted canonical transformations.}
\bea
&&\mbox{\hspace{-19mm}}-\ \sum^{N+1}_{j=1} z_j^* (z_j-z_{j-1})\ - \ i \sum^N_{j=0} H(z_j,z^*_{j+1})\Delta t \nonumber \\
&&\mbox{\hspace{15mm}}= \ F_2(\zeta^*_i,z_i)-\-F_2(\zeta^*_f,z_f) \ + \ \sum^{N+1}_{j=1}\zeta_{j-1}(\zeta^*_j-\zeta^*_{j-1})\non \\
&&\mbox{\hspace{15mm}}- \ i \sum^N_{j=0} H(\zeta_j,\zeta^*_{j+1},\Delta \zeta_j,\Delta \zeta^*_j)\Delta t  \, . \label{exparg}
\eea
Note that the terms in the Hamiltonian depending on $\Delta \zeta_j,\Delta \zeta^*_j$, multiplied by $\Delta t$, always give higher order contributions in the continuous limit and can be thus neglected.

In principle Eq.\eqref{exparg} seems to be the same as the one obtained with more naive approach to the subject. In fact, should we have formally passed to the continuous-time limit and treated $z(t)$ and $\zeta(t)$  as differentiable functions, we would obtain
\be \label{gencont1}
z^*(t) \dot{z}(t) \ = \ \frac{\dr F_2(\zeta^*,z)}{\dr t} \ - \ \zeta (t)\dot{\zeta}^*(t) \, .
\ee
Remembering that
\be
\frac{\dr F_2(\zeta^*,z)}{\dr t}\ = \ \frac{\pa F_2}{\pa \zeta^*}\dot{\zeta}^* \ + \ \frac{\pa F_2}{\pa z}\dot{z}\, ,
\ee
and integrating both members of Eq.\eqref{gencont1} between $t_i$ and $t_f$, we find
\be \label{gencont}
\int^{t_f}_{t_i} z^*(t) \dot{z}(t) \dr t \ = \ F_2(\zeta^*_f,z_f)-F_2(\zeta^*_i,z_i) \ - \ \int^{t_f}_{t_i} \zeta (t) \dot{\zeta}^*(t) \dr t\, .
\ee
Note that Eq.\eqref{gencont} is the continuous version of Eq.\eqref{discgenrel}.
The argument of the PI can be thus written as
\bea \label{parg}
&&\mbox{\hspace{-15mm}}- \int^{t_f}_{t_i} z^*(t)\dot{z}(t) \dr t-i \int^{t_f}_{t_i}  H(z,z^*)\dr t \non \\
&&\mbox{\hspace{-4mm}}= \ F_2(\zeta^*_i,z_i)-F_2(\zeta^*_f,z_f) +\int^{t_f}_{t_i} \zeta(t)\dot{\zeta}^*(t) \dr t-i \int^{t_f}_{t_i}  H(\zeta,\zeta^*)\dr t  \,.
\eea
that is exactly the expression Eq.\eqref{exparg} after that limits $N \rightarrow \infty, \, \epsilon \rightarrow 0$ are taken. However, this kind of argument works only for specific cases and cannot be generally trusted (see discussion in Appendix C).

Coming back to our main considerations, a more elegant form of the kernel Eq.\eqref{initpathint} in the new basis can be obtained noting that
%
\be
F_2(\zeta^*_f,z_f)-F_2(\zeta^*_i,z_i) \ = \ \int^{t_f}_{t_i} \dr t \lf(\dot{\zeta}^*\frac{\pa F_2(\zeta^*,z_i)}{\pa \zeta^*}+\dot{z} \frac{\pa F_2(\zeta^*_f,z)}{\pa z}\ri)\, .
\ee
Now, writing $\dot{z}$ as
\be
\dot{z} \ = \  \frac{\pa z}{\pa \zeta}\dot{\zeta}+\frac{\pa z}{\pa \zeta^*}\dot{\zeta}^* \, ,
\ee
we get that
\begin{eqnarray}
&&\mbox{\hspace{-13mm}}F_2(\zeta^*_f,z_f)-F_2(\zeta^*_i,z_i) \nonumber \\
&&\mbox{\hspace{-5mm}}=\int^{t_f}_{t_i} \dr t \lf[\lf(\frac{\pa F_2(\zeta^*,z_i)}{\pa \zeta^*}+\frac{\pa F_2(\zeta^*_f,z)}{\pa z}\frac{\pa z}{\pa \zeta^*}\ri)\dot{\zeta}^*+\frac{\pa F_2(\zeta^*_f,z)}{\pa z}\frac{\pa z}{\pa \zeta}\dot{\zeta} \ri]  .
\end{eqnarray}
With this we can formally write the PI form of the evolution kernel Eq.\eqref{initpathint} in the new variables as
\bea \label{newker}
\lan z^*_f,t_f|z_i,t_i\ran \ = \ \int^{z^*(t_f)=z^*_f}_{z(t_i)=z_i} \hspace{-0.2 cm}\mathcal{D}\zeta^* \mathcal{D}\zeta \, e^{z^*_f z_f+i\int^{t_f}_{t_i} \dr t \lf[i \lf(Z^*\dot{\zeta}+Z\dot{\zeta}^*\ri)-H(\zeta^*,\zeta)\ri]} \, ,
\eea
where
\be
Z^* \ = \ \lf(\frac{\dr g}{\dr z}\frac{\pa z}{\pa \zeta}\ri) \, ,  \qquad
Z \ = \ \lf(\frac{\dr f}{\dr \zeta^*}+\frac{\dr g}{\dr z}\frac{\pa z}{\pa \zeta^*}-\zeta\ri) \label{phcor2} \, ,
\ee
with $f(\zeta^*)\equiv F_2(\zeta^*,z_i)$ and $g(z)\equiv F_2(\zeta^*_f,z)$.

We stress that the effect of the canonical transformation is to change the Hamiltonian to new one of the form
\be
H(\zeta^*,\zeta)=\frac{\lan\!\lan \zeta^*| \hat{H}\lf(\hat{\al}, \hat{\al}^\dagger\ri)|\zeta\ran\!\ran}{\lan\!\lan \zeta^*|\zeta\ran\!\ran}\, ,
\ee
and a correction to the symplectic phase term appearing in Eq.\eqref{newker}. More subtle effect of this change of representation is reflected in the form of the boundary conditions. We discuss this aspect in detail in Section~\ref{excan2} and in Appendices~A and~B.

We may note in passing that in the context of linear canonical transformations we do not generate any Liouville anomalies. This because the generating function $F_2$ is quadratic and according to Appendix~C it does not produce any non-trivial (anomalous) terms in the Jacobian.

\section{Examples} \label{excan}

We can now apply the techniques developed in the previous section to the van Hove and to Bogoliubov model (cf. also Section~\ref{SEc2}). Since these two systems are described by quadratic Hamiltonians, they can be diagonalized by linear canonical transformations. This simple-looking framework will serve as a convenient testbed allowing us to address a variety of subtle issues involved in canonical transformations within PIs.
In addition, the results obtained will be generalized in Section~\ref{ptqft} to an infinite number of degrees of freedom. This in turn will allow us to see how the inequivalent representations thus emerged
can be interpreted  physically.
To stay as close as possible to QFT (which is our ultimate goal) we will evaluate in this section the (canonically transformed) Euclidean partition functions for these two models.

\subsection{van Hove model} \label{SEc6}

We first consider a single-mode version of the van Hove Hamiltonian Eq.\eqref{VHHam}, i.e.
\be
\hat{H}_{vH} \ = \ \om \hat{a}^\dagger \hat{a} \ + \ \nu(\hat{a}+\hat{a}^\dagger) \, .
\ee
The Hamiltonian $\hat{H}_{vH}$ can be diagonalized by means of a canonical transformation (translation)
\be \label{transop}
\hat{\al} \ = \ \hat{a}\ + \ g \, , \qquad \hat{\al}^\dagger \ = \ \hat{a}^\dagger \ + \ g^*\, ,
\ee
with $g=\nu/\om$. The associated classical transformation are
\be \label{transpoint}
\zeta \ = \ z \ + \ g \, , \qquad \zeta^*\ = \ z^* \ + \ g^*\, .
\ee
The corresponding type-$2$ generating function $F_2$ has the form
\be
F_2(\zeta^*,z) \ = \ z\zeta^* \ - \ g^* z \ + \ g\zeta^* \, . \label{ftwotrans}
\ee
%

Employing the identity Eq.\eqref{idepart}, we can write the evolution kernel in the form
%
%
\bea
\lan z^*_f,t_f|z_i,t_i\ran &=&\lim_{N\rightarrow \infty} \prod^{N}_{j=1} \int \frac{\dr \zeta_j^* \dr \zeta_j}  {2 \pi i}e^{-\sum^{N+1}_{j=1} \zeta_j^* (\zeta_j-\zeta_{j-1})+\zeta^*_f \zeta_f-\frac{\nu}{\omega}(\zeta^*_f +\zeta_i)+\lf(\frac{\nu}{\om}\ri)^2} \non  \\
& \times & e^{-i \sum^N_{j=0}\lf(\om \zeta^*_{j+1}\zeta_j-\frac{\nu^2}{\om}\ri)\Delta t} \,.
\eea
Let us note that because of linearity of this transformations there are no anomalous corrections. We can thus write
\bea
\lan z^*_f,t_f|z_i,t_i\ran =\int \mathcal{D}\zeta^* \mathcal {D}\zeta e^{\zeta^*_f \zeta_f-\frac{\nu}{\omega}(\zeta^*_f +\zeta_i)+\lf(\frac{\nu}{\om}\ri)^2}  e^{-\int^{t_f}_{t_i}  \zeta^* \dot{\zeta}-i \lf(\om \zeta^*\zeta-\frac{\nu^2}{\om}\ri) \dr t} \,.
\eea
This is proportional to the harmonic oscillator kernel. In fact, the extra pieces are fixed by the mixed boundary conditions. This can be exactly solved in the saddle point approximation. We thus get:
\be
\lan z^*_f,t_f|z_i,t_i\ran=e^{\zeta^*_f e^{-i\om(t_f-t_i)}\zeta_i-\frac{\nu}{\omega}(\zeta^*_f +\zeta_i)+\lf(\frac{\nu}{\om}\ri)^2-i (t_f-t_i)\frac{\nu^2}{\om}} \, .
\ee
Passing to the Euclidean regime, the kernel takes the form:
\be \label{kervan}
\lan z^*_f,\beta|z_i,0\ran=e^{\zeta^*_f e^{-\beta \om}\zeta_i-\frac{\nu}{\omega}(\zeta^*_f +\zeta_i)+\lf(\frac{\nu}{\om}\ri)^2-\beta\frac{\nu^2}{\om}} \, .
\ee
The Euclidean partition function
\be
\mathcal{Z} \ = \ \int \dr \mu(z) \lan z^*, \beta | z, 0 \ran \, .
\ee
can be now easily evaluated. With this we finally arrive at the result
\be
\mathcal{Z}_{vH} \ = \ \mathcal{Z}_{ho} \ \! e^{-\beta\frac{\nu^2}{\om}} \, ,
\ee
where
\be
\mathcal{Z}_{ho} \ = \ \frac{1}{1-e^{- \bt \om}}\, ,
\ee
is the partition function of the one-dimensional linear harmonic oscillator~\cite{ItzZub,Kle}.

\subsection{Bogoliubov model} \label{excan2}

In this section we evaluate the Euclidean partition function of a system described by a single-mode Bogoliubov Hamiltonian
\be \label{bog}
\hat{H}_B \ = \ \frac{k^2}{2m}\hat{a}^\dagger \hat{a} \ + \ \frac{N^2}{2V}U_0 \ + \ \frac{N}{2V}U_0 \lf[\hat{a}^{\dagger \, 2}+\hat{a}^2 \ + \ 2 \hat{a}^\dagger \hat{a} \ri] .
\ee

We notice that $\hat{H}_B$ can be diagonalized by the time independent Bogoliubov--Valatin transformation  [cf. Eqs.(\ref{dynmapsupa}),(\ref{dynmapsup})]
\bea
\hat{\al}  &=&  \hat{a} \ \! \cosh \theta   \ + \ \hat{a}^\dagger \ \! \sinh \theta   \, , \\
\hat{\al}^\dagger  &=&  \hat{a} \ \! \sinh \theta  \ + \ \hat{a}^\dagger \ \! \cosh \theta   \, .
\eea
The related classical transformation reads
\bea \label{classbog1}
\zeta &=& z \ \!  \cosh \theta  \ + \ z^* \ \! \sinh \theta   \, , \\
\zeta^* &=&  z \ \! \sinh \theta  \ + \ z^* \ \! \cosh \theta   \, .
\label{classbog2}
\eea
It can be checked that the type-2 generating function for this transformation is
\be \label{genbog}
F_2(\zeta^*,z) \ = \ \ha \tanh \theta\lf(\zeta^{* \, 2}-z^2 \ri) \ + \ z\zeta^* \ \! \mathrm{sech} \ \! \theta \, .
\ee
However when we try to evaluate the kernel along the same lines as for the van Hove model, i.e, by performing above canonical transformation and then using the saddle point approximation, we arrive at seeming difficulty. The transformation Eq.\eqref{classbog1} mixes $z$ and $z^*$ while~\eqref{transpoint} was a point transformation. In classical Hamiltonian dynamics it is required to fix both canonical variables at final and initial times when one performs a canonical transformations, in order to avoid problems arising when integrating by parts~\cite{Goldstein}. This is impossible in quantum mechanics because of the obvious conflict with uncertainty principle. In Eq.\eqref{initpathint} it is clear that this imposition would be an over-specification of the problem. This fact reflects the non commutativity of $\hat{a}$ and $\hat{a}^\dagger$. These considerations let some authors as far as to claim that any canonical transformations in the functional framework are not well defined~\cite{Heg}. Here we prove that this assertion is not fully justified, at least not, in the case of linear transformations.

Let us consider the following function written in terms of new variables:
\bea \label{mats}
i\mathcal{S} &= & z^*_f z_f(\zeta_f,\zeta^*_f) \ + \ i\lf\{i\lf[F_2(\zeta^*_f,z_f)-F_2(\zeta^*_i,z_i)\ri] \ri. \\ \non
&-& \lf. \int^{t_f}_{t_i} i \zeta(t)\dot{\zeta}^*(t) \dr t \ - \ \int^{t_f}_{t_i}  H(\zeta,\zeta^*)\dr t\ri\} \, .
\eea
Integrating by parts we have:
\bea
i\mathcal{S} &=& z^*_f z_f(\zeta_f,\zeta^*_f)\ + \ \zeta^*_f \zeta_f-\zeta^*_i \zeta_i \ - \ \lf[F_2(\zeta^*_f,z_f)\ - \ F_2(\zeta^*_i,z_i)\ri]  \non \\ \label{ac}
 &-&\int^{t_f}_{t_i} \zeta^*(t)\dot{\zeta}(t) \dr t \ - \ i \int^{t_f}_{t_i}  H(\zeta,\zeta^*)\dr t
\eea
The problem under study requires to impose
\be \label{fixedvar}
\delta z^*_f \ = \ \delta z_i \ = \ 0 \, .
\ee
Since the transformation is linear we get Hamilton equations for new variables:
\bea
\dot{\zeta}^*  \ = \ i \frac{\pa H(\zeta, \zeta^*)}{\pa \zeta}  \label{hamnew1}\, , \qquad
\dot{\zeta} \ = \ -i \frac{\pa H(\zeta, \zeta^*)}{\pa \zeta^*} \, .
\eea
However our boundary condition has the form:
\bea
z\lf(\zeta(t_i),\zeta^*(t_i)\ri) \ = \ z_i \, , \qquad z^*\lf(\zeta(t_f),\zeta^*(t_f)\ri) \ = \ z^*_f \, ,
\eea
namely only a combination of new variables can be fixed. Further details are relegated to~\ref{calvar}.

Let us turn back to our specific problem. The Bogoliubov Hamiltonian is reduced to the harmonic oscillator form by using Bogoliubov transformation Eqs.\eqref{classbog1}-\eqref{classbog2}. The Hamilton equations~\eqref{hamnew1} thus become
\bea
\dot{\zeta}^*  \ = \  i \epsilon \,  \zeta^* \, , \qquad
\dot{\zeta}  \ = \ -i \epsilon \,  \zeta \, .
\eea
Their solution is:
\bea
\zeta^*  \ = \ B \, e^{i \epsilon t}  \, , \qquad
\zeta  \ = \  A \, e^{-i \epsilon t} \, .
\eea
where $\ep$ is given by Eq.\eqref{enspequa}  and $A$ and $B$ have to be determined in terms of $z_i$ and $z^*_f$ by the boundary conditions
\bea
z_i &=& \cosh \theta \ \zeta(t_i) \ - \ \sinh \theta \ \zeta^*(t_i) \, , \\
z^*_f &=& -\sinh \theta \ \zeta(t_f) \ + \ \cosh \theta \ \zeta^*(t_f) \, .
\eea
We thus find
\bea \label{coeffsol1}
A &=& \frac{z_i \cosh \theta e^{i \epsilon t_f} \ + \ z^*_f \sinh \theta e^{i \epsilon t_i}}{\cosh^2 \theta e^{i \epsilon (t_f-t_i)} \ - \  \sinh^2 \theta e^{-i \epsilon (t_f-t_i)}} \, , \\
B &=& \frac{z^*_f \cosh \theta e^{-i \epsilon t_i} \ + \ z_i \sinh \theta e^{-i \epsilon t_f}}{\cosh^2 \theta e^{i \epsilon (t_f-t_i)} \ - \ \sinh^2 \theta e^{-i \epsilon (t_f-t_i)}} \label{coeffsol2} \, .
\eea
The kernel is now evaluated thanks to semiclassical approximation Eq.\eqref{semiclassicalsol}. To evaluate the partition function we are interested in the case:
\be
t_f \ = \ T \, , \,\,\, t_i \ = \ 0  \, , \,\,\, z_i \ = \ z \, , \,\,\, z^*_f\ = \ z^* \, .
\ee
After Wick rotating the result to the Euclidean regime, we obtain
\bea
&&\mbox{\hspace{-18mm}} \lan z^*, \beta | z, 0\ran \ = \  e^{-\beta E_0}\sqrt{\frac{\mathrm{sech}^2\theta}{1-e^{-2\epsilon \beta}\tanh^2 \theta}} \non \\
&&\times \ \exp\lf[\frac{\mathrm{sech}^2 \theta  z^* e^{-\epsilon \beta}z+\ha\lf(z^2+z^{* \, 2}\ri)\lf(e^{-2 \beta \epsilon}-1\ri)\tanh \theta}{1-e^{-2\epsilon \beta}\tanh^2 \theta} \ri]  ,
\eea
where $E_0$ has been defined in Eq.\eqref{zeropoint}. The partition function is, finally:
\be
\mathcal{Z}_B \ = \ \frac{e^{-\beta E_0}}{1-e^{-\beta \epsilon}} \ = \  \mathcal{Z}_{ho} \ \! e^{-\beta E_0} \, ,
\ee
which coincides with the result found in Ref.~\cite{GerSil}.

In passing we notice that the application of the semiclassical approximation correctly reproduces the zero point energy contribution of the quasi-particle Hamiltonian. Indeed, from Eq.\eqref{semiclassicalsol} we get, in the present case:
\be
\exp\lf[\frac{i}{2}\frac{\pa^2 H}{\pa \zeta \pa \zeta^*}\lf(\frac{\pa \zeta}{\pa z^*}\frac{\pa \zeta^*}{\pa z}+\frac{\pa \zeta^*}{\pa z^*}\frac{\pa \zeta}{\pa z}\ri)\ri]=e^{\frac{i}{2}(t_f-t_i)\ep}e^{-\frac{i}{2}(t_f-t_i) \lf[\ep-\frac{k^2}{2m}-mu^2\ri]} \, .
\ee
The first piece on the RHS is canceled by the Pauli--van Vleck--Morette like term
\be
\sqrt{i\frac{\pa^2 S}{\pa z^*_f \pa z_i}} \ = \ e^{-\frac{i}{2}(t_f-t_i)\ep} \, .
\ee
In the operatorial approach this contribution would appear because of non-commutativity of $\hat{\al}$ and $\hat{\al}^\dagger$.

\section{Functional integrals and canonical transformations in QFT} \label{SEc7}

Here we treat canonical transformation in the framework of coherent state FIs for field theoretical systems. We shall see how inequivalent representations of CCR affect our study when we pass to the description of systems in an infinite volume.

\subsection{Quantum system in a box}

Here we extend the analysis of Section~\ref{SEc4} to QFT systems. By following Section~\ref{SEc2}, we first analyze systems confined in a box of volume $V$. Moreover, because of freedom in boundary conditions on the box walls (they do not affect the result in the infinite volume limit), we choose Born--von Karman (i.e., periodic) boundary conditions so that momentum $\G k$ can assume only the discrete values $\G k=(2 \pi)^3 \G n / V$ where $\G n=(n_x,n_y,n_z)$ is an integer-valued vector. The large-V limit will be taken at the very end through the relations Eqs.\eqref{ptvo1}-\eqref{ptvo3}.

To proceed, we start with some basic definitions. Let $\{z_\G k\}$ be a sequence of complex numbers so that
\be
\sum_{\G k} z^*_{\G k} z_{\G k}  \ < \ \infty  \, .
\ee
A coherent states is defined as
\be
|\{z\}\ran \ = \ e^{-\frac{1}{\sqrt{V}}\sum_{\G k} z_{\G k} \hat{a}^\dagger_\G k}|0\ran \, .
\ee
These are, once more, eigenstates of annihilation operators:
\be
\hat{a}_{\G k}|\{z\}\ran \ = \ \frac{z_{\G k}}{\sqrt{V}} |\{z\}\ran \,.
\ee
The bra vectors are defined as
\be
\lan \{z^*\}| \ = \ \lan 0| e^{-\frac{1}{\sqrt{V}}\sum_{\G k} z^*_{\G k} \hat{a}_\G k} \, ,
\ee
and are eigenstates of creation operators:
\be
\lan \{z^*\} | \hat{a}^\dagger_{\G k} \ = \  \lan \{z^*\}| \frac{z^*_{\G k}}{\sqrt{V}}  \, .
\ee
The Heisenberg-picture base vectors are  defined as
\be
|\{z\},t\ran \ = \ e^{-i \hat{H}(\{\hat{a}\},\{\hat{a}^\dagger\})t} |\{z\}\ran \, .
\ee
A resolution of the identity can be written in terms of these states as
\be
\int [\dr \mu(z)]|\{z\},t\ran \lan \{z^*\},t| \ = \ \ide \, ,
\ee
where
\be
 [\dr \mu(z)]=\prod_\G k \dr \mu(z_{\G k})=\frac{1}{V}\prod_\G k  \frac{\dr z_{\G k} \dr z^*_{\G k}}{2 \pi i} e^{-\frac{1}{V}\sum_{\G k} z^*_{\G k} z_{\G k}} \, .
\ee
The scalar product between two coherent states is given by:
\be
\lan \{\tilde{z}^*\}|\{z\}\ran \ = \ e^{\frac{1}{V}\sum_{\G k} \tilde{z}_{\G k}^*z_{\G k}} \, .
\ee
Therefore we can construct a time-sliced expression for the evolution kernel in terms of coherent states:
\begin{eqnarray}
&&\mbox{\hspace{-15mm}}\lan \{z^*\}_f,t_f|\{z\}_i,t_i\ran \ = \ \lim_{N\rightarrow \infty} \prod^{N}_{j=1} \int [\dr \mu(z_{j }^*) \dr \mu(z_{j})] e^{\frac{1}{V} \sum_\G k z^*_{f \,\G k} z_{f \, \G k}} \nonumber
 \\
&&\mbox{\hspace{5mm}}\times \ e^{i\sum^{N+1}_{j=1}\sum_{\G k} \lf[\frac{i}{V} \, z^*_{j \, \G k}(z_{j \, \G k}-z_{j-1 \, \G k})\ri]-i\lf[\sum^{N}_{j=0} H(\lf\{z \ri\}_{j+1},\lf\{z^*\ri\}_j)\Delta t\ri]} \, . \label{feynkera}
\end{eqnarray}
Here
\be
H(\lf\{z \ri\}_{j+1},\lf\{z^*\ri\}_j)\  \equiv \ \frac{\lan \{z^*\}_{j+1},t_{j+1}|\hat{H}(\lf\{\hat{a}\ri\},\lf\{\hat{a}^\dagger\ri\})|\{z\}_j, t_j \ran}{\lan \{z^*\}_{j+1},t_{j+1}|\{z\}_j, t_j \ran} \, .
\ee
Eq.(\ref{feynkera}) can be in the continuous limit formally written as
\bea
\lan \{z^*\}_f,t_f|\{z\}_i,t_i\ran &=& \int^{\{z^*\}(t_f)=\{z^*\}_f}_{\{z\}(t_i)=\{z\}_i} \prod_\G k\mathcal{D}z^*_{\G k} \mathcal{D}z_{\G k} e^{\frac{1}{V}\sum_{\G k} z^*_{f \, \G k} z_{f \, \G k}} \nonumber \\
& \times &e^{i\int^{t_f}_{t_i} \dr t \lf[\frac{i}{V}\sum_{\G k} z^*_{\G k}(t) \dot{z}_\G k(t)- H(\lf\{z \ri\},\lf\{z^*\ri\})\ri]} \, .
\label{initpathinta}
\eea
Following Section \ref{SEc4}, we may now consider another set of coherent states $|\{\zeta\}\ran\!\ran$ which span the Hilbert space $\tilde{\mathcal{H}}$:
\bea
|\{\zeta\} \ran\!\ran&=& e^{\frac{1}{\sqrt{V}}\sum_{\G k} \zeta_{\G k} \hat{\al}^\dagger_\G k}|0\ran\!\ran\, , \qquad
\lan \! \lan {} \{\zeta^*\}| \ = \ \lan\!\lan 0|e^{\frac{1}{\sqrt{V}}\sum_{\G k} \zeta^*_{\G k} \hat{\al}_\G k} \, ,
\\
\hat{\al}_{\G k} |\{\zeta\}\ran\!\ran\, &=& \,\frac{\zeta_{\G k}}{\sqrt{V}} |\{\zeta\}\ran\!\ran\, , \mbox{\hspace{10mm}} \lan\!\lan \{\zeta^*\}| \hat{\al}^\dag_\G k \,=\, \frac{\zeta^*_{\G k}}{\sqrt{V}}\lan \!\lan \{\zeta^*\}| \, ,
\eea
where
\be
\hat{\al}_\G k \ = \ \exp(-i \hat{\mathcal{K}}) \, \hat{a_\G k} \, \exp(i \mathcal{K})\, , \qquad \hat{\al}^\dagger_\G k \ = \ \exp(-i \mathcal{K}) \,  \hat{a}^\dagger_{\G k} \, \exp(i \hat{\mathcal{K}})\, .
\ee
The self-adjoint operator $\hat{\mathcal{K}}$ has the form [cf. Eq.(\ref{66a})]
\be
\hat{\mathcal{K}} \ = \ \sum_{\G k} \theta_\G k \hat{K}(\hat{a}_\G k,\hat{a}^\dagger_\G k) \, .
\ee
Mixed scalar products can be represented through FIs as
\bea
 \lan \{z^*\} |\{\zeta\}\ran\!\ran &=& \int^{\{\xi^*\}(\theta)=\{z^*\}}_{\{\xi\}(0)=\{\zeta\}} \prod_\G k\mathcal{D}\xi^*_\G k \mathcal{D}\xi_\G k e^{\frac{1}{V}\sum_{\G k} \xi^*_\G k (\theta_k) \xi_\G k(0)}  \non \\
& \times &  e^{i\int^{\theta_\G k}_{0} \dr \theta'_\G k [\frac{i}{V}\sum_{\G k}\xi^*_\G k(\theta'_\G k) \dot{\xi}_\G k(\theta'_\G k)-\sum_{\G k}K(\xi_{\G k}^*,\xi_\G k)]} \, , \label{funqft1} \\[3mm] \non
\lan\!\lan \{\zeta^*\} |\{z\}\ran &=& \int^{\{\xi^*\}(\theta)=\{\zeta^*\}}_{\{\xi\}(0)=\{z\}} \prod_\G k\mathcal{D}\xi^*_\G k \mathcal{D}\xi_\G k e^{\frac{1}{V}\sum_{\G k} \xi^*_\G k (\theta_\G k) \xi_\G k(0)}\\
& \times & e^{i\int^{\theta_\G k}_{0} \dr \theta'_\G k [\frac{i}{V}\sum_{\G k} \xi^*_\G k (\theta'_\G k) \dot{\xi}_\G k(\theta'_\G k)-\sum_{\G k} K(\xi^*_{\G k},\xi_\G k)]} \, . \label{funqft2}
\eea
In analogy with Section~\ref{SEc4} we can write the solutions for the above mixed scalar products as
\bea
\lan \{z^*\} |\{\zeta\}\ran\!\ran&=& A^*\lf(\{\zeta^*\},\{z\}\ri)\exp\lf\{F^*\lf(\{\zeta\}^*,\{z\}\ri)\ri\} \label{fanansatz1} \, ,\\[1mm]
\lan\!\lan \{\zeta^*\} |\{z\}\ran &=& A\lf(\{\zeta^*\},\{z\}\ri)\exp\lf\{F\lf(\{\zeta\}^*,\{z\}\ri)\ri\} \, . \label{fanansatz2}
\eea
Functions $A$ and $F$ can be evaluated in the WKB approximation, which for linear canonical transformations yields exact results, namely
\bea \label{mmsa1}
\hspace{-0.5 cm} A\lf(\{\zeta^*\},\{z\}\ri) &=&\prod_\G k \lf(\frac{\pa^2 F_2\lf(\{\zeta^*\},\{z\}\ri)}{\pa z_{\G k} \pa \zeta^*_{\G k}}\ri)^\ha \,, \\ \label{mmsa2}
\hspace{-0.5 cm} F\lf(\{\zeta^*\},\{z\}\ri) &=& F_2\lf(\{\zeta^*\},\{z\}\ri)+\sum_{\G k}\frac{i}{2} \int^{\theta_\G k}_{0} \dr \theta'_\G k \frac{\pa^2 K(\xi^c_\G k,\xi^{* \, c}_\G k)}{\pa \xi_\G k \pa \xi^*_\G k} \,  .
\eea
Here
\be
F_2\lf(\{\zeta\}^*,\{z\}\ri) \ = \ \frac{1}{V}\sum_{\G k} F_{2 \, \G k}(\zeta^*_{\G k},z_{\G k}) \, ,
\ee
is the classical type-2 generating function of the canonical transformation and $\xi_\G k^c, \xi^{* \, c}_\G k$ are solution of the equations
\bea
\frac{\dr \xi_\G k}{\dr \theta} = -i \frac{\pa K(\xi^{*}_\G k,\xi_\G k)}{\pa \xi^*_\G k} \, , \qquad
\frac{\dr \xi^{*}_\G k}{\dr \theta} = i \frac{\pa K(\xi^{*}_\G k,\xi_\G k)}{\pa \xi_\G k} \,  .
\eea
For a finite $V$ the above two representations are always unitarily equivalent. In fact, employing the same argument as in Section~4, we again obtain that
\be
\frac{\lan\!\lan \{\zeta^*\}_{j+1} | \{z\}_{j+1} \ran}{\lan\!\lan \{\zeta\}^*_{j+1} | \{z\}_j \ran} \ = \ e^{\mathcal{G}(\{\zeta\}^*_{j+1},\{z\}_{j+1})-\mathcal{G}(\{\zeta\}^*_{j+1},\{z\}_j)}  \, ,
\label{165bb}
\ee
with
\begin{eqnarray}
\mathcal{G}\lf(\{\zeta\}^*,\{z\}\ri) \ = \ F\lf(\{\zeta\}^*,\{z\}\ri)+\log A\lf(\{\zeta\}^*,\{z\}\ri)\, ,
\end{eqnarray}
is well defined. If we now multiply both sides of Eq.(\ref{165bb}) by $\lan \{z\}^*_{j+1} | \{\zeta\}_{j+1} \ran$ $\times \lan \{\zeta\}^*_{j+1} | \{z\}_j \ran$, and integrate over $\lf[\dr \mu(\zeta_{j+1})\ri]$ we get
\bea
\lan \{z^*\}_{j+1} | \{z\}_{j+1} \ran &=& \int \lf[\dr \mu(\zeta_{j+1})\ri] \lf[e^{\mathcal{G}(\{\zeta\}^*_{j+1},\{z\}_{j+1})-\mathcal{G}(\{\zeta\}^*_{j+1},\{z\}_j)} \ri. \non  \\
& \times & \lf. \lan \{z\}^*_{j+1} | \{\zeta\}_{j+1} \ran\!\ran\lan\!\lan \{\zeta\}^*_{j+1} | \{z\}_j \ran \ri] \, .
\eea
With this we arrive at the consistency relation [cf. Eq.\eqref{dscthirdgen1}]
\be \label{cons1}
\mathcal{G}(\{\zeta\}^*_{j+1},\{z\}_{j+1})-\mathcal{G}(\{\zeta\}^*_{j+1},\{z\}_j) \ = \  \frac{1}{V}\sum_{\G k} z_{j+1 \, \G k}^* (z_{j+1 \, \G k}-z_{j \, \G k})   \, .
\ee
Similarly we find [cf. Eq.\eqref{dscthirdgen2}] that
\be \label{cons2}
\mathcal{G}(\{\zeta\}^*_{j+1},\{z\}_{j})-\mathcal{G}(\{\zeta\}^*_{j},\{z\}_j) \ = \ \frac{1}{V} \sum_{\G k} \zeta_{j \, \G k} (\zeta^*_{j+1 \, \G k}-\zeta^*_{j \, \G k}) \, .
\ee
Following Section~\ref{SEc4} we can replace  in above expressions $\mathcal{G}\lf(\{\zeta\}^*,\{z\}\ri)$ with $F_2\lf(\{\zeta\}^*,\{z\}\ri)$. As a result, we find that the argument of the FI in new variables reads
\bea
&& \mbox{\hspace{-9mm}} -\sum^{N+1}_{j=1}\sum_{\G k} \lf[\frac{1}{V} \, z^*_{j \, \G k}(z_{j \, \G k}-z_{j-1 \, \G k})\ri]-i\lf[\sum^{N}_{j=0} H(\lf\{z \ri\}_{j+1},\lf\{z^*\ri\}_j)\Delta t\ri] \non \\
&&  \mbox{\hspace{-9mm}} = \ \non F_2(\{\zeta\}^*_i,\{z\}_i)-F_2(\{\zeta\}^*_f,\{z\}_f)+\frac{1}{V}\sum^{N+1}_{j=1} \sum_{\G k} \zeta_{j-1 \, \G k} (\zeta^*_{j \, \G k}-\zeta^*_{j-1 \, \G k})\\
&& \mbox{\hspace{-9mm}} - \ i \sum^N_{j=0} H(\lf\{\zeta_{j}\ri\},\lf\{\zeta^*_{j+1}\ri\},\lf\{\Delta \zeta_{j}\ri\},\lf\{\Delta \zeta^*_{j}\ri\})\Delta t  \label{exparginf}  \,.
\eea
A continuous-time limit expression for the FI in the new basis can be thus formally written as
\bea \label{newkerinf}
\lan \{z\}^*_f,t_f|\{z\}_i,t_i\ran &=& \int^{\{z^*\}(t_f)=\{z^*\}_f}_{\{z\}(t_i)=\{z\}_i} \prod_\G k\mathcal{D}\zeta^*_{\G k} \mathcal{D}\zeta_{\G k} e^{\frac{1}{V}\sum_{\G k} z^*_{f \, \G k} z_{f \, \G k}}\\ \non
&\times & e^{i\int^{t_f}_{t_i} \dr t \lf[\frac{i}{V} \sum_{\G k} \lf( Z^*_{\G k}\dot{\zeta}_k+Z_{\G k}\dot{\zeta}^*_k\ri)-H(\lf\{\zeta\ri\},\lf\{\zeta^*\ri\})\ri]} \, ,
\eea
where
\begin{eqnarray}
Z_{\G k} \ = \ \lf(\frac{\dr f_\G k}{\dr \zeta^*_{\G k}}+\frac{\dr g_\G k}{\dr z_{\G k}}\frac{\pa z_{\G k}}{\pa \zeta^*_{\G k}}-\zeta_{\G k}\ri) ,  \mbox{\hspace{10mm}}
Z^*_{\G k} \ = \ \lf(\frac{\dr g_\G k}{\dr z_{\G k}}\frac{\pa z_{\G k}}{\pa \zeta_{\G k}}\ri)\, ,
\label{usefsym2}
\end{eqnarray}
and
\be
 f_\G k(\zeta^*_{\G k}) \ \equiv \ F_{2 \, \G k}(\zeta^*_{\G k},z_{\G k}) \, , \qquad  g_\G k(z_{\G k})\ \equiv \ F_{2 \, \G k}(\zeta^*_{f \, \G k},z_{\G k}) \, .
\ee
In the following section we will discuss  what happens when we remove the finite-$V$ regulator.

\subsection{Infinite volume limit and inequivalent representations} \label{ptqft}

As seen in Section~\ref{SEc2}, quantum systems with infinite number of degrees of freedom differ from finite (i.e., first-quantized) ones in the crucial respect that the algebraic relations governing their observables generally admit inequivalent irreducible representations. This is typically epitomized by the fact that in the QFT limit the Stone--von Neumann theorem cannot be applied, and we must deal with many
unitarily inequivalent ground states and ensuing Hilbert spaces. This, in turn, provides a conceptual platform allowing to discuss topics (such as continuous phase transitions~\cite{UMZ2,Sewell,BJV} or renormalization~\cite{BJV,UTK,Haag}) that are otherwise inaccessible in
the realm of QM.

Let us first note that in the large $V$ limit Eqs.\eqref{funqft1}-\eqref{funqft2} [or equivalently Eqs.\eqref{fanansatz1}-\eqref{fanansatz2}] acquire the form\footnote{By analogy with Section~\ref{vhop} we define $z(\G k)=z_\G k (2 \pi)^{3/2}$ and $z^*(\G k)=z^*_\G k (2 \pi)^{3/2}$.}
\bea
\lan \lf\{z^*\ri\} | \lf\{\zeta\ri\}\ran\!\ran&=& \exp\lf\{\intk \ F^*_2\lf[\zeta^*(\G k),z(\G k)\ri]+C^*\ri\} \label{fanansatz3} \, ,\\
\lan\!\lan \lf\{\zeta^*\ri\} | \lf\{z\ri\}\ran &=& \exp\lf\{\intk \ F_2\lf[\zeta^*(\G k),z(\G k)\ri]+C\ri\} \, , \label{fanansatz4}
\eea
where $C$ is a constant given by
\begin{eqnarray}
C &=& \frac{V}{8 \pi^3}\intk \, \lf\{\ha\log i\frac{\de^2 F_2\lf[\zeta^*(\G k),z(\G k)\ri]}{\de \zeta^*(\G k) \de z(\G k)}\right. \nonumber \\[2mm]
&&+\ \left. \frac{i}{2} \int^{\theta}_{0} \dr \theta' \frac{\de^2 K\lf[\xi^*(\G k),\xi(\G k)\ri]}{\de \xi(\G k) \de \xi^*(\G k)}\ri\} \ + \ C_u \, .
\label{ourformulatocont}
\end{eqnarray}
Here $C_u$ is a classical contribution resulting from the generating functional $F_2$. Classically this is typically fixed by requiring that no explicit time-dependent terms are present in the generating function (as we work with restricted canonical transformations)~\cite{FaMa}. Since $C_u$  is not explicitly time dependent, it is typically set to zero in classical considerations. However, in quantum theory $C_u$ cannot be automatically set to zero because its value is fixed by the vacuum structure. In fact, we may notice that from Eq.\eqref{fanansatz4} one has
\be
\lan\!\lan 0|0 \ran \ = \  \exp(C) \, .
\ee
Value of $C$ can be explicitly determined by using the functional representations Eqs.\eqref{funqft1} and \eqref{funqft2} for mixed scalar products. By imposing the boundary conditions $\zeta(\G k, \theta'_\G k=0)=\zeta^*(\G k, \theta'_\G k=\theta_\G k)=0$ we get
\be \label{Intgen1}
\lan\!\lan 0|0 \ran \ = \ \prod_\G k\int_{\xi(0)=0}^{\xi^*(\theta_\G k)=0} \mathcal{D} \xi(\G k)  \mathcal{D} \xi^*(\G k) e^{i\intk \ W\lf[\zeta(\G k),\zeta^*(\G k)\ri]} \, ,
\ee
with the Lagrange-like density term
\be
W\ = \ \int^{\theta_\G k}_0 \dr \theta' \lf[i\xi^*(\G k,\theta'_\G k) \frac{\dr \xi(\G k, \theta'_\G k)}{\dr \theta'_\G k}-K[\xi(\G k,\theta'_\G k),\xi^*(\G k,\theta'_\G k)]\ri] \, .
\label{177a}
\ee
The FI described by Eq.\eqref{Intgen1} can be explicitly determined through the WKB approximation. Taking into account that we have, in this case, the classical identification
\be \label{regens}
W \lf[\zeta^*(\G k), z(\G k)\ri] \ = \ F_2\lf[\zeta^*(\G k),z(\G k)\ri] \, ,
\ee
we obtain
\be
C_u \ = \ \intk \ W \lf[\zeta^c(\G k),\zeta^{* \, c}(\G k)\ri] \, ,
\ee
where $\zeta^c({\G k},\theta)$ and $\zeta^{* \, c}({\G k},\theta)$ satisfy the classical-like equations
\bea
\frac{\de W \lf[\zeta(\G k),\zeta^{* }(\G k)\ri]}{\de \zeta({\G k},\theta)} \ = \ 0 \, , \qquad
\frac{\de W \lf[\zeta(\G k),\zeta^{* }(\G k)\ri]}{\de \zeta^{*}({\G k},\theta)} \ = \ 0 \,  ,
\eea
with boundary conditions $\zeta(\G k, \theta'_\G k=0)=\zeta^*(\G k, \theta'_\G k=\theta_\G k)=0$. There are no additional constants. In fact, one can check that in the particular case when $K=0$, we correctly obtain $\lan 0| 0\ran=1$. This condition, which is of a pure quantum nature, removes the classical ambiguity in $C$.

We will now use  our two previous examples to show how these generic considerations work in practice. For simplicity's sake, we drop the dependence on $\G k$ while doing the calculation. \\

\noindent i) In the case of {\em translations} [see Eq.\eqref{transop}] we get
\be
K(\xi^*,\xi) \ = \ i\lf(\xi-\xi^*\ri) \, ,
\ee
which gives
\be
\lan\!\lan 0|0 \ran \ = \ \int_{\xi(0)=0}^{\xi^*(g)=0} \mathcal{D} \xi  \ \! \mathcal{D} \xi^* \ \! e^{i\int^g_0 \dr g' \lf[i\xi^* \frac{\dr \xi}{\dr g'} \ + \ i\xi^*\ - \ i\xi\ri]} \, .
\ee
Eq.\eqref{pseduhameq2} appears as
\bea
\frac{\dr \xi}{\dr g'} \ = \ -1 \, , \qquad
\frac{\dr \xi^*}{\dr g'} \ = \ -1 \, ,
\eea
with solution
\bea
\xi \ = \ -g' \, , \qquad
\xi^* \ = \ -g'+g \, .
\eea
With these we get [see Eq.\eqref{sempropagator1}]
\be
\lan\!\lan 0|0 \ran \ = \ \exp\!\left[-\int^g_0 \dr g' g'\right] \ = \ \exp\lf(-\frac{g^2}{2}\ri) .
\ee
By restoring the $\G k$-dependence we can write that
\be
\lan\!\lan 0|0 \ran \ = \  \exp\!\left[-\intk \int^{g_\G k}_0 \dr g'_\G k \ \! g'_\G k\right] \ = \ \exp\lf(-\intk \ \! \frac{g_{\G k}^2}{2}\ri) \, ,
\ee
in accord with Eq.\eqref{vvvH}. Therefore the constant $C$ assumes the value
\be
C \ = \ C_u \ = \-\intk \ \! \frac{g_{\G k}^2}{2} \, . \label{CeqCu}
\ee
Note that the first two contributions to $C$ are null.\\

\noindent ii) In the case of {\em Bogoliubov--Valatin transformations}
\be
K(\xi^*,\xi)\ = \ \frac{i}{2}\lf(\xi^2-\xi^{*\, 2}\ri) \, .
\ee
Eq.\eqref{pseduhameq2} turns out to be
\bea
\frac{\dr \xi}{\dr \theta'}\ = \ -\xi^* \, , \qquad
\frac{\dr \xi^*}{\dr \theta'} \ = \  \xi \, .
\eea
The only solution of these two equations, with boundary conditions $\xi(\theta)=\xi(0)=0$ is trivial, namely  $\xi(\theta') = \xi^*(\theta')  = 0$. So, the sole  contribution to the final result comes from the first term in the sum on the RHS of Eq.\eqref{ourformulatocont}, i.e.
\be
\lan\!\lan 0|0 \ran\ = \ \exp\!\left[\ha \log i\frac{\pa^2 F_2\lf(\zeta^*, z\ri)}{\pa z \pa \zeta^*}\right] \ = \ \exp\!\left[-\ha\log(\cosh \theta)\right],
\label{VI.223.a}
\ee
where we have used Eq.\eqref{genbog}.  In fact, since the complex logarithm is not a single-valued function,
the result Eq.(\ref{VI.223.a}) is true only up to a phase factor. Extending Eq.(\ref{VI.223.a}) to the field-theoretical
case we can employ the box regularization and write
\be
\lan\!\lan 0|0 \ran \ = \ \exp\!\left[-\sum_\G k \log(\cosh \theta_\G k)\right] ,
\ee
that in the large volume limit coincides with Eq.\eqref{vvB}. These results are in agreement with results, derived in a different way in Ref.~\cite{Berezin}.

If the above two representations are unitarily inequivalent, Hilbert spaces are orthogonal and the vacuum-to-vacuum amplitude Eq.\eqref{Intgen1} is zero. It is clear from our examples that this divergence must come from the constant $C$ alone. Therefore we cannot employ naively the same reasonings that allowed us to arrive at Eqs.\eqref{cons1}-\eqref{cons2}. Consequently, we should understand that the ensuing continuous-time  FI
\begin{eqnarray}
&&\mbox{\hspace{-15mm}}\lan \lf\{z^*\ri\}_f,t_f| \lf\{z_i\ri\},t_i\ran \ = \  \prod_{\G k} \int^{\tilde{z}^*(t_f)=z^*_{f}}_{\tilde{z}(t_i)=z_i} \mathcal{D}\zeta^* (\G k)\mathcal{D}\zeta^*(\G k) \ \! e^{\intk \ \tilde{z}^*_f \tilde{z}_f} \nonumber \\[1mm]
&&\mbox{\hspace{15mm}}\times \ e^{i \intk\lf\{i\lf(F_2[\zeta^*_f(\G k),\tilde{z}_f]\ - \ F_2[\zeta_i^*(\G k),\tilde{z}_i]\ri)\ri\}}\nonumber \\[1mm] &&\mbox{\hspace{15mm}}\times \ e^{-i\int^{t_f}_{t_i} \dr t \ \lf[i \zeta(\G k,t) \dot{\zeta}^*(\G k,t)\ + \ H(\zeta(\G k,t),\zeta^*(\G k,t))\ri]} \, ,
\label{pathgenbas}
\end{eqnarray}
is only a formal object whose meaning is provided only via  regulating scheme.
In Eq.(\ref{pathgenbas}) we indicated $\tilde{z}=z\lf(\zeta(\G k,t),\zeta^*(\G k,t)\ri)$, as a formal limit
of the Eq.\eqref{newkerinf}. By employing the classical relation Eq.\eqref{regens} we may rewrite this kernel in a more suggestive way, namely
\begin{eqnarray}
&&\mbox{\hspace{-15mm}} \lan \lf\{z^*\ri\}_f,t_f| \lf\{z_i\ri\},t_i\ran \ = \  \prod_{\G k} \int^{\tilde{z}^*(t_f)=z^*_{f}}_{\tilde{z}(t_i)=z_i} \mathcal{D}\zeta^* (\G k)\mathcal{D}\zeta^*(\G k) \ \! e^{\intk \ z^*_f z_f} \nonumber \\[1mm]
&&\mbox{\hspace{15mm}}\times \ e^{i \intk\lf\{i\lf(W[\zeta^*_f(\G k),\tilde{z}_f] \ - \ W[\zeta_i^*(\G k),\tilde{z}_i]\ri)\ri\}}\nonumber \\[1mm]
&&\mbox{\hspace{15mm}}\times \ e^{-i \int^{t_f}_{t_i} \dr t \ \lf[i \zeta(\G k,t) \dot{\zeta}^*(\G k,t) \ + \  H(\zeta(\G k,t),\zeta^*(\G k,t))\ri]} \, ,
\label{pathgenbas1}
\end{eqnarray}
We note in particular that the quantum generator of the canonical transformation now appears in the exponent. If two representations of CCR are unitarily inequivalent, the unitary generator of the canonical transformation does not exist and then also the classical generating function of the corresponding classical linear transformation is badly defined\footnote{In fact, it contains $C_u$ that  can be a divergent quantity as in Eq.\eqref{CeqCu} when $g_{\G k} \notin L^2({\G {R}}^{\!3})$.}. For this reason, we stressed that these above FI representations are only formal. Strictly speaking, one should perform all computations in the box regularization (as done in the previous section) and only at the very end of the computations the large-$V$ limit should be taken.
In the rest of this section we will illustrate the physical meaning of inequivalent representations by using the correspondence between QFT and statistical mechanics.

Let us first consider the partition function for the van Hove model. Repeating the same steps as in Section~\ref{SEc6} we find that the Euclidean kernel of the van Hove model reads
\begin{eqnarray} \label{kervaninf}
&&\mbox{\hspace{-15mm}}\lan \{z^*\}_f,\beta|\{z\}_i,0\ran \nonumber \\
&&\mbox{\hspace{-11mm}} = \ \exp\left\{\frac{1}{V}\sum_{\G k} \lf[\zeta^*_{f \, \G k} e^{-\beta \om_{\G k}}\zeta_{i \, \G k}-\frac{\nu_{\G k}}{\omega_{\G k}}(\zeta^*_{f \, \G k} +\zeta_{i \, \G k})+\lf(\frac{\nu_{\G k}}{\om_{\G k}}\ri)^2-\beta\frac{\nu_{\G k}^2}{\om_{\G k}}\ri]\right\} .
\end{eqnarray}
With this the ensuing partition function acquires the form
\be
\mathcal{Z}_{vH}\ = \ \mathcal{Z}_{nBg}\ \! e^{-\frac{\beta}{V} \sum_{\G k}\frac{\nu_{\G k}^2}{\om_{\G k}}} \, ,
\label{227a}
\ee
where
\be
\mathcal{Z}_{nBg} \ = \ \exp\!\lf[-\sum_{\G k} \log\lf(1-e^{-\beta \om_{\G k}}\ri)\ri] \! ,
\ee
represents the partition function of a non-interacting  Bose gas of (spinless) quasi-particles~\cite{LiPi}.
To understand better the physical meaning of the obtained results, we evaluate the ensuing Helmholtz free energy $\mathcal{F} = -(1/\beta)\log \mathcal{Z} $. The energy gap between these two systems can be found by looking at
\be
\Delta {\mathcal{F}}^{(1)} \ \equiv \  {\mathcal{F}}_{vH}\ - \ {\mathcal{F}}_{nBg} \ = \ \frac{1}{V}\sum_\G k\frac{\nu_{\G k}^2}{\om_\G k} \, ,
\ee
In case when we consider the translationally invariant van Hove model [i.e. when $\nu(\G k)/\om(\G k)= c \delta(\G k)$] we obtain in the large-$V$ limit
\be
\Delta {\mathcal{F}}^{(1)} \ = \ c^2\int \mathrm{d}^3 {\G k} \ \om (\G k)\delta^2(\G k)\ = \ \frac{V c^2 \om(\G 0)}{(2 \pi)^3} \, .
\label{sec.6.230a}
\ee
So, in particular, in the thermodynamical limit the RHS of Eq.(\ref{sec.6.230a}), i.e. the energy difference between the two representations, diverges. That is, in nutshell, the physical meaning of the unitary inequivalence in QFT: an infinite energy is required to pass from one vacuum state to another one. Note also, that the difference of free energy densities, i.e.
\be
\Delta {\Phi}^{(1)} \ \equiv \ {\Phi}_{vH}\ - \ {\Phi}_{nBg} \ = \ \frac{c^2}{V} \int \mathrm{d}^3 {\G k} \  \om (\G k)\delta^2(\G k)\ = \ \frac{c^2 \om(\G 0)}{(2 \pi)^3} \, ,
\ee
remains finite as expected. In the same way, we find for the Bogoliubov model the energy gap $\Delta {\mathcal{F}}^{(2)} \equiv {\mathcal{F}}_{B}\ - \ {\mathcal{F}}_{nBg}$ of the form
\be
\Delta {\mathcal{F}}^{(2)}  \ = \ \ha N m u^2 +\frac{V}{2(2 \pi)^3} \int \mathrm{d}^3 {\G k} \lf[\ep(\G k)-\frac{{\G k}^2}{2m}-mu^2+\frac{m^3 u^4}{{\G k}^2}\ri]\! ,
\ee
i.e. the energy gap is again divergent. However the energy density gap
\be
\Delta {\Phi}^{(2)} \ = \ \ha  m n u^2 +\frac{1}{2(2 \pi)^3} \int \mathrm{d}^3 {\G k} \lf[\ep(\G k)-\frac{{\G k}^2}{2m}-mu^2+\frac{m^3 u^4}{{\G k}^2}\ri] \! ,
\label{232bb}
\ee
(here $n=N/V$) remains finite. It is interesting to extend the thermodynamic analysis of Eqs.(\ref{sec.6.230a})-(\ref{232bb}) a bit further.
By using the Maxwell relations
\begin{eqnarray}
\left(\frac{\partial \mathcal{F}}{\partial \Theta} \right)_{V,N} \ = \ - S, \;\;\;\; \; \left(\frac{\partial \mathcal{F}}{\partial  V} \right)_{ \Theta,N} \ = \ -p\, ,
\end{eqnarray}
where $\Theta = 1/\beta$, $S$ and $p$ are the temperature, entropy and pressure of a system, respectively, we see that
\begin{eqnarray}
\left(\frac{\partial \Delta\mathcal{F}^{(1,2)}}{\partial \Theta} \right)_{V,N} \ = \ -\Delta S \ = \ 0, \;\;\;\;\;\;  \left(\frac{\partial \Delta\mathcal{F}^{(1,2)}}{\partial V} \right)_{\Theta,N} \ = \ -\Delta p \ = \ 0\, .
\end{eqnarray}
It should be stressed that in order to obtain these results one should, according to standard rules of statistical physics, perform first the indicated derivatives when $V$ is finite and take the thermodynamic limit only at the very end.

Above behavior is typical for continuous phase transitions often encountered in QFT in connection with spontaneous symmetry breaking. In fact,  continuous phase transition can be understood as a passage between unitarily inequivalent Hilbert spaces which represent physically distinct thermodynamical phases, see, e.g., Refs.~\cite{UMZ2,BJV,BJ33,V}.

\section{Conclusions and Outlook} \label{SEc8}

In this paper we have analyzed  the r\^{o}le of unitarily inequivalent representations of CCR  in QFT from the functional-integral standpoint.
%
%
To better understand the issue at stake we carried out our discussion in the framework of canonical transformations which represent a paradigmatic mechanism for generation of inequivalent representations in QFT. This was illustrated with two emblematic examples; the van Hove model and the Bogoliubov model of a weakly interacting Bose gas. We have found that the information about a particular representation is reflected in FIs
in the form of the transformed Hamiltonian, in the appearance of the quantum generator $W$ and in the form of the boundary conditions~[cf. Eq.(\ref{pathgenbas1})].


The issue of canonical transformations within path-integral formalism has been discussed by a number of authors~\cite{Fan,GeJe,Schu1982,FuKa,Shi}, exhibiting many subtle points especially in connection with non-linear canonical transformations. Many of the reported difficulties found in these approaches could be traced down to an uncritical use of generic canonical transformations and to the phase-space formulation of PIs. As pointed out by~\cite{KlaSka,Klauder}, some of the difficulties related to use of the phase-space PI can be ameliorated in the coherent-state PIs. On the other hand, by confining ourselves to linear canonical transformations (which are, due to the Groenewold--van Hove theorem, the only ones that are unitarily implementable in QM) we obtained a sound mathematical basis for our PI analysis. So, in order to incorporate the canonical transformations in both the path and functional-integral formalism we worked with the coherent-state PIs and FIs. Apart from the aforementioned mathematical convenience (e.g., better behaved measure) this was also dictated by applications in QFT where one works with creation and annihilation operators which are directly translatable into complex variables in PIs and FIs. In fact, Glauber coherent FIs are presently indispensable as a model-building tool in a number of condensed-matter systems~\cite{Miransky}.

It seems to be a common wisdom among path-integral practitioners that functional integrals are somehow ``representation independent".
Actually, this is not the case as we have seen here. In particular, we have explicitly shown that when performing a linear canonical transformation, we need to rephrase FI in the new representation: in the known cases of canonical transformations giving rise to inequivalent representations, we were able to recover with FIs the same results as with the conventional operatorial approach (e.g., orthogonality of associated Hilbert). In addition, by considering the partition function built on the two inequivalent representations we obtained an infinite energy gap in the thermodynamic limit. This is in a general agreement with the fact that important phenomena such as continuous phase transitions are based on an existence of inequivalent representations of CCR in QFT.

Let us finally add three comments.
%
%
First, in the case when only the partition function is of the interest then the specific representation is imprinted only in the form of the new action and in the quantum generator $W$. As noted in Section~6, the generator $W$ is infinite when the volume regulator is removed. $W$ is thus reminiscent of a  counterterm in QFT which arises when passing from bare to renormalized quantities (such as fields, masses and couplings) or when passing between quantities defined at different renormalization points. This analogy is further reinforced by the fact that Fock spaces defined
at different renormalization points are typically unitarily inequivalent~\cite{BJV,UTK,Haag}.

Second, results contained in the present paper represent a first step in the FI treatment of physical systems characterized by the presence of  inequivalent representations. For example, in the problem of flavor mixing, it has been known for some time that different sets of Green's functions built on two inequivalent vacua (mass and flavor vacuum) exist. Although such vacuum expectation values can be calculated within operatorial methods it is not yet clear how to do this in the FI framework. If this task is accomplished one can use the usual machinery of FI (e.g., effective action approach) to get the phenomenologically relevant gap equations  for the dynamical generation of flavor mixing~\cite{DynMix}.

We finally note that inequivalent representations have been used also to treat quantum fields in the presence of topologically non trivial extended objects, arising as inhomogeneous condensates of quanta \cite{UMZ2,BJV,kinks}. An extension of the FI formalism developed in this paper to these situations is desirable and it is being currently investigated.


\section*{Acknowledgments}
It is pleasure to acknowledge helpful conversations with
G.~Vitiello and J.~Klauder. P.J.  was  supported  by
the Czech  Science  Foundation Grant No. 17-33812L

\appendix
\section{Phase transformations}\label{SEc5}

For completeness' sake, we discuss in this appendix  phase transformations
\be
\hat{\al} \ = \ e^{i\theta} \hat{a} \, , \qquad \hat{\al}^\dagger \ = \ e^{-i\theta} \hat{a} \, .
\label{A.1.a}
\ee
The corresponding (asymptotic-state) vacua are defined as
\be
\hat{a}|0\ran \ = \ 0 \, , \qquad \hat{\al} |0(\theta)\ran \ = \ 0\, ,
\ee
and are related by a one-parameter group transformation
\be
|0(\theta)\ran \ = \ \hat{G}^{-1}|0\ran \, , \qquad \hat{G}\ = \ \exp(i\theta \ \! \hat{a}^\dagger \hat{a}) \, .
\ee
However, it is clear that the two vacua coincide. This can be seen from the fact that
\be
\hat{\al} |0\ran \ = \ e^{i \theta} \hat{a} |0\ran \ = \ 0 \, .
\ee

This conclusion can be seen also on the level of PIs. To this end we consider the holomorphic-representation analogues of Eq.(\ref{A.1.a})
\be
\zeta \ = \ e^{i\theta} z \, , \qquad \zeta^* \ = \ e^{-i\theta} z^* \, . \label{classpahse}
\ee
whose type-2 generating function is
\be
F_2(\zeta^*,z)\ = \ \zeta^* z e^{i\theta} \, . \label{ftwophase}
\ee

It is not difficult to see that a generic evolution kernel Eq.\eqref{initpathint} is invariant under the transformation Eq.\eqref{classpahse}, apart from boundary conditions\footnote{This just reflects the fact that, although we expand the kernel in another representation, the final and the initial states have to remain unchanged.}. Note that, because the generating function Eq.\eqref{ftwophase} is linear in both variables, there are no anomalous terms present (no Jacobi anomaly and no Edwards--Gulyaev anomaly). The PI in the new variables thus reads
\bea
\lan z^*_f,t_f|z_i,t_i\ran &=& \lim_{N\rightarrow \infty} \prod^{N}_{j=1} \int \frac{\dr \zeta_j^* \dr \zeta_j}{2 \pi i}e^{-F_2(\zeta^*_f,z_f)+F_2(\zeta^*_i,z_i)+\zeta^*_f \zeta_f} \non \\
&& \times \  e^{\sum^{N+1}_{j=1}\zeta_{j-1}(\zeta^*_j-\zeta^*_{j-1})-i \sum^N_{j=0} H(\zeta_j, \zeta^*_{j+1})\Delta t} \,.
\eea
Using Eq.\eqref{ftwophase}, and the identity
\be \label{idepart}
\sum^{N+1}_{j=1}\zeta_{j-1}(\zeta^*_j-\zeta^*_{j-1}) \ = \ -\sum^{N+1}_{j=1} \zeta_j^* (\zeta_j-\zeta_{j-1}) \ + \ \zeta^*_f \zeta_f \ - \ \zeta^*_i \zeta_i \, ,
\ee
(i.e. a discrete analogue of the integration by parts) we find that
\begin{eqnarray}
\lan z^*_f,t_f|z_i,t_i\ran  &=&  \lim_{N\rightarrow \infty} \prod^{N}_{j=1} \int \frac{\dr \zeta_j^* \dr \zeta_j}{2 \pi i}e^{-\sum^{N+1}_{j=1} \zeta_j^* (\zeta_j-\zeta_{j-1})+\zeta^*_f \zeta_f} \nonumber \\
 && \times \ e^{-i \sum^N_{j=0} H(\zeta_j, \zeta^*_{j+1})\Delta t} \,.
\end{eqnarray}
In the continuum limit we might thus write
\be
\lan z^*_f,t_f|z_i,t_i\ran \ = \ \int^{\zeta^*(t_f)=e^{-i\theta} z^*_f}_{\zeta(t_i)=e^{i\theta}z_i} \mathcal{D}\zeta^* \mathcal{D} \ \! \zeta e^{\zeta^*_f \zeta_f}e^{i\int^{t_f}_{t_i} \dr t [i \zeta^*(t) \dot{\zeta}(t)-H(\zeta^*,\zeta)]} \, .
\ee
Here only the boundary conditions reveal  the representations in which calculations are done.

We wish now to show that phase transformations do not produce inequivalent representations, with functional techniques. By using the FI representation of the vacuum-vacuum amplitude Eq.\eqref{Intgen1}, we have:
\be \label{qgpt}
K(\xi,\xi^*) \ = \ -\xi^* \xi \theta \, .
\ee
Eq.\eqref{pseduhameq2} has (similarly as in the case of Bogoliubov transformations from Section~\ref{ptqft})  null solution when $\xi(0)=\xi^*(\theta)=0$. Therefore
\be
\lan\!\lan 0|0 \ran=\lf(\frac{\pa^2 F_2\lf(\zeta^*, z\ri)}{\pa z \pa \zeta^*}\ri)^\ha \exp\lf[\frac{\pa^2 K(\xi^c,\xi^{* \, c})}{\pa \xi \pa \xi^*}\ri] \, .
\ee
Using Eq.\eqref{ftwophase} and Eq.\eqref{qgpt} one might easily verify that
\be
\lf(\frac{\pa^2 F_2\lf(\zeta^*, z\ri)}{\pa z \pa \zeta^*}\ri)^\ha \ = \ e^{i\frac{\theta}{2}} \, , \qquad \exp\lf[\frac{\pa^2 K(\xi^c,\xi^{* \, c})}{\pa \xi \pa \xi^*}\ri] \ = \ e^{-i\frac{\theta}{2}} \, ,
\ee
so that $\lan\!\lan 0|0 \ran=1$. Extending this result to the field-theoretical setting  does not yield any new modifications. In other words, these two representations are always unitarily equivalent.

\section{Semiclassical solution in terms of new variables} \label{calvar}

In this appendix we derive two key results that are employed in Section~\ref{excan2}. First of all, we find out the form of the semiclassical approximation for the transition amplitude after a canonical transformation.
Then we discuss the problem of boundary conditions in transformations that mix both $z$ and $z^*$, and show that for linear transformations (if second-order corrections in $\Delta z$ and $\Delta z^*$ can be neglected in Eqs.\eqref{tayleq},\eqref{tayleq2} for $\Delta t \rightarrow 0$), new variables obey Hamilton equations even though we cannot fix both variables at initial and final time.

To start, let us consider the variation of the following functional
\be
iS_r \ = \ -F_2(\zeta^*_f,z_f)+F_2(\zeta^*_i,z_i)-\zeta^*_i \zeta_i +z^*_f z_f +\zeta^*_f \zeta_f\, ,
\ee
which represents a part of the functional $i\mathcal{S}$ introduced in Eq.\eqref{ac}. Since our original problem must satisfy boundary conditions Eq.\eqref{fixedvar}, we can write
\begin{eqnarray}
\de(iS_r) &=& -\ \delta F_2(\zeta^*_f,z_f) \ + \ \delta F_2(\zeta^*_i,z_i) \ - \ \delta \zeta^*_i \zeta_i \ - \  \zeta^*_i \delta\zeta_i\nonumber \\[2mm] &+& z^*_f \delta z_f \ + \ \zeta^*_f \delta \zeta_f \ + \ \zeta_f \delta \zeta^*_f \, .
\end{eqnarray}
This can be further expanded as
\bea
\de(iS_r)&=&-\frac{\pa F_2(\zeta^*_f,z_f)}{\pa \zeta^*_f}\delta \zeta^*_f\ - \ \frac{\pa F_2(\zeta^*_f,z_f)}{\pa z_f}\delta z_f\ + \ \frac{ \pa F_2(\zeta^*_i,z_i)}{\pa \zeta^*_i}\delta \zeta^*_i  \non \\
&-& \delta \zeta^*_i \zeta_i \ - \ \zeta^*_i \delta\zeta^i \ + \ z^*_f \delta z_f \ + \ \zeta^*_f \delta \zeta_f \ +\ \zeta_f \delta \zeta^*_f \, .
\eea
Now we can employ Eqs.\eqref{tayleq},\eqref{tayleq2} to obtain
\be
\de(iS_r) \ = \ \zeta_f^* \delta \zeta_f\ - \ \zeta^*_i \delta \zeta_i \, .
\ee
These two pieces have the opposite sign of the pieces coming out from the integration by part of the variation of the canonical one form
\be
-\int^{t_f}_{t_i} \zeta^*(t)\delta \dot{\zeta}(t)\ = \ -\zeta_f^* \delta \zeta_f \ + \ \zeta^*_i \delta \zeta_i \ + \ \int^{t_f}_{t_i} \dot{\zeta}^*(t) \delta\zeta(t) \, .
\ee
Therefore, the stationary phase approximation leads to the usual Hamilton equations for new variables, i.e.~Eq.\eqref{hamnew1}.

This result ensures that we can use the semiclassical approximation in the usual way. However we must determine how semiclassical approximation formula can be written in terms of new canonical variables. As carefully evaluated in Ref.~\cite{BdeKKS}, the semiclassical propagator can be written as:
\be  \label{sempropagator}
\lan z^*_f ,t_f| z_i ,t_i\ran \ = \ \sqrt{i\frac{\pa^2 S}{\pa z^*_f \pa z_i}}e^{i \mathcal{S}(z^c,z^{* \, c})}\exp\lf(\frac{i}{2} \int^{t_f}_{t_i} \dr t \frac{\pa^2 H(z^c,z^{* \, c})}{\pa z \pa z^*}\ri) \, ,
\ee
where $z^c, z^{* \, c}$ are solutions of the classical equations of motion and $i \mathcal{S}$ was introduced in Eq.\eqref{mats}. Let us now consider the transformation
\bea
z \ \rightarrow \ z(\zeta,\zeta^*)\, , \qquad z^* \ \rightarrow \ z^*(\zeta,\zeta^*) \, .
\eea
Then
\bea
\frac{\pa}{\pa z}\ = \ \frac{\pa \zeta}{\pa z}\frac{\pa }{\pa \zeta}+\frac{\pa \zeta^*}{\pa z}\frac{\pa}{\pa \zeta^*} \, ,\qquad
\frac{\pa}{\pa z^*} \ = \ \frac{\pa \zeta}{\pa z^*}\frac{\pa }{\pa \zeta}+\frac{\pa \zeta^*}{\pa z^*}\frac{\pa}{\pa \zeta^*} \, ,
\eea
and hence
\begin{eqnarray}
\frac{\pa^2 H}{\pa z \pa z^*} & = & \frac{\pa^2 H}{\pa \zeta^2}\frac{\pa \zeta}{\pa z^*}\frac{\pa \zeta}{\pa z}\ + \ \frac{\pa^2 H}{\pa \zeta \pa \zeta^*}\frac{\pa \zeta}{\pa z^*}\frac{\pa \zeta^*}{\pa z}\nonumber \\[2mm]
& + &  \frac{\pa^2 H}{\pa \zeta^{* \, 2}}\frac{\pa \zeta^*}{\pa z^*}\frac{\pa \zeta^*}{\pa z}
\ + \ \frac{\pa^2 H}{\pa \zeta \pa \zeta^*}\frac{\pa \zeta^*}{\pa z^*}\frac{\pa \zeta}{\pa z} \, .
\end{eqnarray}
In particular, if the transformations reduce the Hamiltonian to the linear harmonic oscillator form, then the former reduces to
\be
\frac{\pa^2 H}{\pa z \pa z^*}=\frac{\pa^2 H}{\pa \zeta \pa \zeta^*}\lf(\frac{\pa \zeta}{\pa z^*}\frac{\pa \zeta^*}{\pa z}+\frac{\pa \zeta^*}{\pa z^*}\frac{\pa \zeta}{\pa z}\ri) \, .
\ee
Consequently, the semiclassical propagator written in terms of these new variables reads
\be \label{semiclassicalsol}
\lan z^*_f ,t_f| z_i ,t_i\ran\ = \ \sqrt{i\frac{\pa^2 S}{\pa z^*_f \pa z_i}}\ \! e^{i \mathcal{S}(\zeta^c,\zeta^{* \, c})+\frac{i}{2} \int^{t_f}_{t_i} \dr t \frac{\pa^2 H(\zeta^c,\zeta^{* \, c})}{\pa \zeta \pa \zeta^*}\lf(\frac{\pa \zeta}{\pa z^*}\frac{\pa \zeta^*}{\pa z}+\frac{\pa \zeta^*}{\pa z^*}\frac{\pa \zeta}{\pa z}\ri)} \, .
\ee
%
\section{Coherent state PI and FI and non-linear canonical transformations}

For completeness' sake, we briefly discuss in this Appendix some issues related non-linear canonical transformations in PIs. These cannot be analyzed, in principle, with the same methods employed in Sections~\ref{SEc4} and \ref{SEc7} because of the Groenewold--van Hove theorem. Nevertheless, we show that some interesting results can still be obtained when a less rigorous approach is employed.

By emulating Ref.~\cite{FuKa} we might define a quantum canonical transformation via relations
\bea
F_2(\zeta^*_j,z_j)-F_2(\zeta^*_j,z_{j-1}) &=& z_j^* (z_j-z_{j-1}) \, , \label{discinftrans3} \\
F_2(\zeta^*_j,z_{j-1})-F_2(\zeta^*_{j-1},z_{j-1}) &=& \zeta_{j-1}(\zeta^*_j-\zeta^*_{j-1}) \, , \label{discinftrans4}
\eea
where $F_2(\zeta^*,z)$ is the type-$2$ generating function of a (in principle non-linear) classical canonical transformation,
as in Eqs.(\ref{discinftrans1}),(\ref{discinftrans}).
Now we Taylor-expand these as in Eqs.\eqref{tayleq},\eqref{tayleq2}. So, that we can write
\bea
\label{nltayleq}
z^*_j &=& \frac{\pa F_2(\zeta^*_j,z_j)}{\pa z_j}+\ha\frac{\pa^2 F_2(\zeta^*_j,z_j)}{\pa z^2_j}(z_{j-1}-z_j)+\ldots \, ,  \\
\zeta_j &=&\frac{\pa F_2(\zeta^*_{j},z_j)}{\pa \zeta^*_{j}}+\ha\frac{\pa^2 F_2(\zeta^*_{j},z_j)}{\pa \zeta^{* \, 2}_{j}}(\zeta^*_{j+1}-\zeta^*_j)+\ldots \, ,\label{nlnltayleq2}
\eea
where the dots indicate that, in principle we could have non trivial contributions coming from higher-order terms. Note in this connection that new variables do not depends only on the old ones, but also on their variations $\Delta z_j \equiv z_j-z_{j-1} , \, \Delta z^*_j \equiv z_{j+1}-z_j$. As discussed in Section~2, this can lead to Liouville anomalies. The latter do not exist for linear canonical transformations.

To see this explicitly, let us write the inverse Jacobian of our transformation
\be \label{jacobian}
J^{-1}\ = \ \prod^N_{j=1} \lf[\frac{\pa \zeta_j}{\pa z_j} \frac{\pa \zeta^*_j}{\pa z^*_j}-\frac{\pa \zeta^*_j}{\pa z_j} \frac{\pa \zeta_j}{\pa z^*_j}\ri] \, .
\ee
From Eq.\eqref{nltayleq} we find that:
\bea
\hspace{-0.2 cm}\frac{\pa \zeta_j}{\pa z_j}&=&\frac{\pa^2 F_2(\zeta^*_j.z_j)}{\pa \zeta^*_j \pa z_j}+\ha \frac{\pa^2 F_2(\zeta^*_j.z_j)}{\pa \zeta^{* \, 2}_j}\frac{\pa \zeta^*_j}{\pa z_j}+\ha \frac{\pa^3 F_2(\zeta^*_j.z_j)}{\pa \zeta^{* \, 2}_j \pa z_j} \Delta \zeta^*_j \, ,\non \\
\hspace{-0.2 cm}\frac{\pa \zeta_j}{\pa z^*_j}&=& \ha \frac{\pa^2 F_2(\zeta^*_j.z_j)}{\pa \zeta^{* \, 2}_j}\frac{\pa \zeta^*_j}{\pa z^*_j} \, .
\eea
Therefore, substituting in the Jacobian~\eqref{jacobian} we obtain
\be \label{jacobian1}
J^{-1} \ = \ \prod^N_{j=1} \lf[\lf(\frac{\pa^2 F_2(\zeta^*_j.z_j)}{\pa \zeta^*_j \pa z_j}+\ha \frac{\pa^3 F_2(\zeta^*_j.z_j)}{\pa \zeta^{* \, 2}_j \pa z_j} \Delta \zeta^*_j\ri) \frac{\pa \zeta^*_j}{\pa z^*_j}\ri] \, .
\ee
Noting that
\be
1 \ = \ \frac{\pa z^*_j}{\pa z^*_j}=\frac{\pa^2 F_2(\zeta^*_j.z_j)}{\pa \zeta^*_j \pa z_j}\frac{\pa \zeta^*_j}{\pa z^*_j}-\ha \frac{\pa^3 F_2(\zeta^*_j.z_j)}{\pa z^2_j \pa \zeta^*_j}\frac{\pa \zeta^*_j}{\pa z^*_j}\Delta z_j \, ,
\ee

we thus find
\bea \label{jacobian2}
J^{-1} \ = \ \prod^N_{j=1} \lf[1+\ha \frac{\pa^3 F_2(\zeta^*_j.z_j)}{\pa z^2_j \pa \zeta^*_j}\frac{\pa \zeta^*_j}{\pa z^*_j}\Delta z_j+\ha \frac{\pa^3 F_2(\zeta^*_j.z_j)}{\pa \zeta^{* \, 2}_j \pa z_j} \frac{\pa \zeta^*_j}{\pa z^*_j}\Delta \zeta^*_j \ri] \, .
\eea
Finally, using that
\be
\Delta z_j \ = \ \frac{\pa z_j}{\pa \zeta_j}\Delta \zeta_j+\frac{\pa z_j}{\pa \zeta^*_j}\Delta \zeta^*_j \, ,
\ee
we can write the inverse Jacobian in the form
\be \label{jacobian3}
J^{-1} \ = \ \prod^N_{j=1} \lf[1+A_j \Delta \zeta_j+B_J \Delta\zeta^*_j \ri] \, ,
\ee
where
\bea
A_j &=& \ha \frac{\pa^3 F_2(\zeta^*_j.z_j)}{\pa z^2_j \pa \zeta^*_j}\frac{\pa \zeta^*_j}{\pa z^*_j}\frac{\pa z_j}{\pa \zeta_j} \, , \label{ancor1} \\ \label{ancor2}
B_j &=& \ha\lf(\frac{\pa^3 F_2(\zeta^*_j.z_j)}{\pa z^2_j \pa \zeta^*_j}\frac{\pa \zeta^*_j}{\pa z^*_j}\frac{\pa z_j}{\pa \zeta^*_j}+ \frac{\pa^3 F_2(\zeta^*_j.z_j)}{\pa \zeta^{* \, 2}_j \pa z_j} \frac{\pa \zeta^*_j}{\pa z^*_j}\ri) \, ,
\eea
which corresponds to the Eq.\eqref{jacps} in the phase-space framework. By exponentiating the inverse Jacobian as
\be \label{jacexp}
J^{-1} \ = \ e^{\log \lf[\prod^N_{j=1}\lf(1+A_j \Delta \zeta_j+B_j \Delta \zeta^*_j\ri)\ri]}\ \sim \  e^{\sum^N_{j=1}\lf( A \Delta \zeta_j+B \Delta \zeta^*_j \ri)} \, ,
\ee
Eq.\eqref{jacexp} shows that, if $\Delta \zeta_j$ and $\Delta \zeta^*_j$  are of order $O(\Delta t)$ (ballistic regime), they can give a non-trivial contribution to the final result. In fact, in that case
\be \label{jacexpt}
J^{-1} \ = \ e^{\sum^N_{j=1}\Delta t\lf( A_j \frac{\Delta \zeta_j}{\Delta t}+B_j \frac{\Delta \zeta^*_j}{\Delta t} \ri)} \, ,
\ee
which, in the limit $N \rightarrow \infty$ (or equivalently $\Delta t \rightarrow 0$) can be rewritten as
\be \label{jacexptfunc}
J^{-1} \ = \  e^{\int^{t_f}_{t_i} \dr t \lf( A(t) \dot{\zeta}(t)+B(t) \dot{\zeta}^*(t) \ri)} \, .
\ee
In QFT case, the Jacobian takes the form
\bea \label{jacobianinf}
J^{-1} = \prod^N_{j=1} \prod_\G k \lf[1+A_{j \, \G k} \Delta \zeta_{j \, \G k}+B_{j \, \G k}\Delta \zeta^*_{j \, \G k} \ri] \, ,
\eea
where
\bea
A_{j \, \G k} &=& \ha \frac{\pa^3 F_{2 \, \G k}(\zeta^*_{j \, \G k}.z_{j \, \G k})}{\pa z^2_{j \, \G k} \pa \zeta^*_{j \, \G k}}\frac{\pa \zeta^*_{j \, \G k}}{\pa z^*_{j \, \G k}}\frac{\pa z_{j \, \G k}}{\pa \zeta_{j \, \G k}} \, , \\
B_{j \, \G k} &=& \ha\lf(\frac{\pa^3 F_{2 \, \G k}(\zeta^*_{j \, \G k}.z_{j \, \G k})}{\pa z^2_{j \, \G k} \pa \zeta^*_{j \, \G k}}\frac{\pa \zeta^*_{j \, \G k}}{\pa z^*_{j \, \G k}}\frac{\pa z_{j \, \G k}}{\pa \zeta^*_{j \, \G k}}+ \frac{\pa^3 F_{2 \, \G k}(\zeta^*_{j \, \G k}.z_{j \, \G k})}{\pa \zeta^{* \, 2}_{j \, \G k} \pa z_{j \, \G k}} \frac{\pa \zeta^*_{j \, \G k}}{\pa z^*_{j \, \G k}}\ri) \, . \\
\eea
We can thus rewrite the evolution kernel along the same lines as in Eq.\eqref{newkerinf}:
\bea
\lan \{z\}^*_f,t_f|\{z\}_i,t_i\ran &=& \int^{\{z^*\}(t_f)=\{z^*\}_f}_{\{z\}(t_i)=\{z\}_i} \prod_\G k\mathcal{D}\zeta^*_{\G k} \mathcal{D}\zeta_{\G k} e^{\frac{1}{V}\sum_{\G k} z^*_{f \, \G k} z_{f \, \G k}}\\ \non
&\times & e^{i\int^{t_f}_{t_i} \dr t \lf\{\frac{i}{V} \sum_{\G k} \lf[ Z^*_{\G k}\dot{\zeta}_k+Z_{\G k}\dot{\zeta}^*_k\ri]-H(\lf\{\zeta\ri\},\lf\{\zeta^*\ri\})\ri\}} \, ,
\eea
but now
\bea
Z_{\G k} &=& \lf(\frac{\dr f_\G k}{\dr \zeta^*_{\G k}}+\frac{\dr g_\G k}{\dr z_{\G k}}\frac{\pa z_{\G k}}{\pa \zeta^*_{\G k}}-\zeta_{\G k}-B_\G k V\ri) ,  \\
Z^*_{\G k} &=& \lf(\frac{\dr g_\G k}{\dr z_{\G k}}\frac{\pa z_{\G k}}{\pa \zeta_{\G k}}-A_\G k V\ri),
\eea
i.e. anomalous corrections contribute to the symplectic phase term.




\begin{thebibliography}{99}
\section*{References}

\bibitem{schweber}
S.~Schweber, {\it An Introduction to Relativistic Quantum Field Theory}, (Row, Peterson and Co.,London, 1962).

\bibitem{ItzZub}
C.~Itzykson and J.B.~Zuber, {\it Quantum Field Theory}, (McGraw-Hill, Inc., New York, 1980).

\bibitem{Neu} J.~von Neumann, Ann.\ Math. {\bf 104} (1931)  570.

\bibitem{bratt} see, e.g., O.~Bratteli  and  D.~Robinson, {\it Operator  Algebras  and Quantum   Statistical   Mechanics}, (Springer,  New  York,
1979).

\bibitem{Kastrup} H.A.~Kastrup, Fortsch. \ Phys. {\bf 51} (2003) 975.

\bibitem{UMZ2} H.~Umezawa, {\it Advanced field theory: Micro, Macro, and Thermal Physics}, (AIP, New York, 1993).

\bibitem{Sewell} G.~Sewell, {\it Quantum Mechanics and Its Emergent Macrophysics}, (Princeton University Press, Oxford, 2002).
	
\bibitem{BJV} M.~Blasone, P.~Jizba and G.~Vitiello, {\it Quantum Field Theory and its Macroscopic Manifestations},
(World Scientific, London, 2011).

\bibitem{UTK} H.~Umezawa, Y.~Takahashi and S.~Kamefuchi, Ann. \ Phys. {\bf 26} (1964) 336.

\bibitem{kinks}
  M.~Blasone and P.~Jizba,
  Ann. \ Phys.\  {\bf 295} (2002) 230.

\bibitem{Haag} R.~Haag, {\it Local Quantum Physics: Fields, Particles, Algebras}, (Springer, Berlin, 1996).

\bibitem{Haag2} R.~Haag, H.~Narnhofer and U.~Stein, Commun. \ Math.  \ Phys. {\bf 94} (1984) 219.

\bibitem{Hawking}
S.W.~Hawking, Commun. Math. Phys. {\bf 43} (1975)  199.

\bibitem{MaSoVi}
M.~Martellini, P.~Sodano and G.~Vitiello, Il Nuovo Cimento {\bf 48} (1978) 341.

\bibitem{Birrell}
 N.~D.~Birrell and P.~C.~W.~Davies,
 {\it Quantum Fields in Curved Space} (Cambridge U. Press, Cambridge, 1984).

\bibitem{CeRaVi}
E.~Celeghini, M.~Rasetti and G.~Vitiello, Ann. \ Phys. {\bf 215} (1992) 156.

\bibitem{Mixing}
M.~Blasone and G.~Vitiello,   Ann. Phys. (N.Y.) (1995) {\bf 244} 283;
M.~Blasone, A.~Capolupo, O.~Romei and G.~Vitiello,
Phys.\ Rev.\ D {\bf 63} (2001) 125015;
M.~Blasone, A.~Capolupo and G.~Vitiello,
Phys.\ Rev.\ D {\bf 66} (2002) 025033;
M.~Blasone, M.V.~Gargiulo and G.~Vitiello,
Phys. \ Lett. \ B {\bf 761} (2016)  104-110.


\bibitem{BHV99}
M.~Blasone, P.A.~Henning and G.~Vitiello,
Phys.\ Lett.\ B {\bf 451} (1999) 140;
 M.~Blasone, P.~Pires Pacheco and H.~W.~C.~Tseung,
  Phys.\ Rev.\ D {\bf 67} (2003) 073011.

\bibitem{DynMix}
M.~Blasone, P.~Jizba, G.~Lambiase and N.E.~Mavromatos,
J.\ Phys. {\bf 538} (2014)  012003;
M.~Blasone, P.~Jizba and L.~Smaldone, Il Nuovo Cimento {\bf 38 C} (2015) 201.





\bibitem{Blasone:2015bea}
M.~Blasone, G.~Lambiase and G.G.~Luciano,
J. \ Phys. \ Conf. \ Ser.  {\bf 631}  (2015) 012053.
	
\bibitem{MatPapUme}
H.~Matsumoto, N.J.~Papastamatiou and H.~Umezawa, Phys.\ Lett. {\bf 46}{\rm B} (1973)  73.
	
\bibitem{Torre}
C.G.~Torre, Phys.\ Rev. D {\bf 72} (2005)  025004.

\bibitem{TelNog}
A.~Teleki and M.~Noga, eprint arXiv:hep-th/0601136 (2006).

\bibitem{Groe} H.J.~Groenewold,  Physics {\bf 12} (1946) 405.

\bibitem{VH1951}
L.~van Hove, Mem.\  Academ. \ Roy. \ Belg. {\bf 26} (1951)  610.

\bibitem{Swa}
M.S.~Swanson, Phys.\ Rev. A {\bf 50} (1994)  4538.	

\bibitem{Umezawa:1982nv}
H.~Umezawa, H.~Matsumoto and M.~Tachiki,
{\it Thermo Field Dynamics And Condensed States}, (North-Holland, Amsterdam, 1982).

\bibitem{FadSlav}
L.D.~Faddeev and A.A.~Slavnov, {\it Gauge Fields-Introduction to Quantum Theory}, (The Benjamin Cummings Publishing Company, New York, 1980).

\bibitem{And}
A.~Anderson, Ann.\  Phys. {\bf 232} (1994) 292-331.
	
\bibitem{BJSPrague}
M.~Blasone, P.~Jizba and L.~Smaldone, to be published in J.\ Phys. \ Conf. \ Ser. (2017).

\bibitem{vanHove}
L.~van Hove, Physica {\bf 18} (1952)  145.

\bibitem{Bog}
N.~Bogoliubov, J.\  Phys. \ USSR {\bf 11} (1947)  23.
		
	
\bibitem{Tung}
W.K.~Tung, {\it Group Theory in Physics}, (World Scientific, Singapore, 1980).

\bibitem{Miransky}
V.A.~Miransky, {\it Dynamical Symmetry Breaking in Quantum Field Theories}, (World Scientific, London, 1993).
	
\bibitem{LiPi}
E.M.~Lifshitz and L.P.~Pitaevskii, {\it Statistical Physics, Vol.2}, (Pergamon Press, Oxford, 1980).

\bibitem{StWi}
R.F.~Streater and A.S.~Wightman, {\it PCT, Spin and Statistics and all that}, (W.A.Benjamin, New York, 1964).
	
\bibitem{Fan}
R.~Fanelli, J. \ Math.\ Phys. {\bf 17} (1976) 490.

\bibitem{GeJe}
J.L.~Gervais and A.~Jevicki, Nucl. \ Phys. B {\bf 110} (1976)  93.

\bibitem{Schu1982}
L.S.~Schulman, {\it Techniques and Applications of Path Integration}, (John Wiley and Sons, New York, 1982).

\bibitem{FuKa}
H.~Fukutaka and T.~Kashiwa, Ann.\ Phys. {\bf 185} (1988)  301.

\bibitem{Shi}
A.Y.~Shiekh, J.\ Math. \ Phys. {\bf 36} (1995)  6681.

\bibitem{Kle}
H.~Kleinert, {\it Path Integrals in Quantum Mechanics, Statistics, Polymer Physics, and Financial Markets}, (World Scientific, London, 2009).

\bibitem{Goldstein}
H.~Goldstein, C.~Poole and J.~Safko, {\it Classical Mechanics}, (Addison Wesley, London, 1950).

\bibitem{EG} S.F.~Edwards and Y.V.~Gulyaev, Proc. \ R. \ Soc. \ A {\bf 279} (1964) 229.

\bibitem{SchuMc}
D.W.~McLaughlin nad L.~S.~Schulman, J.\ Math. \ Phys. {\bf 12} (1971)  2520.


\bibitem{Schweber}
S.S.~Schweber, J.\ Math. \ Phys. {\bf 3} (1962)  831.

\bibitem{KlaSka}
J.R.~Klauder and B.~Skagerstam , {\it Coherent States-Applications in Physics and in Mathematical Physics}, (World Scientific, New York, 1985).
	
\bibitem{Klauder}
J.R.~Klauder, {\it A Modern Approach to Functional Integration}, (Birkh\"auser, London, 2011).

\bibitem{Swabook}
M.S.~Swanson, {\it Path Integrals and Quantum Processes}, (Academic Press, San Diego, 1992).
		
\bibitem{Perelomov}
A.~Perelomov, {\it Generalized Coherent states and Their Applications}, (Springer-Verlag, Berlin, 1986).
		
\bibitem{Fan1975}
R.~Fanelli, J. \ Math. \ Phys. {\bf 16} (1975) 1729.

\bibitem{Strocchi}
F.~Strocchi, Rev. \ Mod. \ Phys. {\bf 38} (1966)  36.

\bibitem{Testa1970}
F.J.~Testa, J.\ Math. \ Phys. {\bf 11} (1970)  2698.
	
\bibitem{Testa1973}
F.J.~Testa, Physica {\bf 69} (1952)  579.

\bibitem{BdeKKS}
M.~Baranger, M.~A.~M.~de Aguiar, F.~Keck, H.~J.~Korsch and B.~Schellaa, J.\ Phys. \ A {\bf 34} (2001)  7227.
	
\bibitem{Wei}
Y.~Weissman, Journ. \ Chem. \ Phys. {\bf 76} (1982) 4067.

\bibitem{Hel1977}
E.J.~Heller, Journ. \ Chem. \ Phys. {\bf 66} (1977) 5777.

\bibitem{Heg}
J.~Hegseth, eprint arXiv:quant-ph/0403005 (2004).
	
\bibitem{GerSil}
C.C.~Gerry and S.~Silverman, J.\ Math. \ Phys. {\bf 23} (1982)  1995.
	
\bibitem{FaMa}
A.~Fasano, S.~Marmi, {\it Analytical Mechanics}, (Oxford graduate texts, Oxford, 2006).

\bibitem{Berezin}
F.A.~Berezin, {\it The Method of Second Quantization}, (Academic Press Inc., London, 1966).

\bibitem{BJ33} M.~Blasone and P.~Jizba, J. Phys. A: Math. Theor. {\bf 45} (2012)  244009.

\bibitem{V} G.~Vitiello, Int. \ J.\  Mod.  \ Phys. \ B {\bf 18} (2004) 705.

\bibitem{private communication}
J.R.~Klauder, private communication.




	
\end{thebibliography}
\end{document}